\documentclass[reprint,nofootinbib,amsmath,amssymb,aps,floatfix]{revtex4-2}
\usepackage{graphicx}
\usepackage{dcolumn}
\usepackage[colorlinks,linkcolor=magenta,anchorcolor=cyan,citecolor=blue]{hyperref}
\usepackage{bm}
\usepackage[multiple]{footmisc}
\usepackage{appendix}

\begin{document}
\title{Superradiance and quasinormal modes of the gravitational perturbation around rotating hairy black hole}
\author{Zhen Li}
\email{zhen.li@nbi.ku.dk}
\affiliation {DARK, Niels Bohr Institute, University of Copenhagen, Jagtvej 128, 2200 Copenhagen Ø, Denmark}
\date{\today}

\begin{abstract}
The No Hair theorem in classical general relativity predicts that rotating black holes are specified by the Kerr metric, which is uniquely identified by the mass and spin. However, as a pioneering study beyond general relativity, the rotating hairy black hole has been proposed, which encompasses the Kerr black hole as a special case. In these black holes, there are extra hair which could appear due to the
 additional surrounding sources such as dark matter or dark energy. In this work, we study the phenomenology of the rotating hairy black hole in terms of gravitational perturbations. In particular, the supperradiance and the quasinormal modes. Using the matching-asymptotic method, we derive the amplification factor and the superradiance conditions. We also calculate the quasinormal modes using the continued fraction method. The results are in very good agreement with previous studies in the Kerr limit. We also show how the amplification and quasinormal modes will shift in response to variations in the hairy parameters, black hole spin, and quantum numbers.
\end{abstract}
\maketitle

\section{introduction}

Recent observations of gravitational waves (GW) \cite{gw1,gw2,gw3} and black hole shadows \cite{shadow1,shadow2} provide strong evidence for the existence of black holes, the intriguing objects predicted by general relativity. It is widely believed that the physically rotating black holes in the universe are Kerr black holes described by the Kerr metric. The famous No Hair theorem \cite{nh1,nh2,nh3,nh4,nh5} also states that the Kerr metric is completely determined by the mass and spin of the black hole. However, due to the additional surrounding sources like dark matter, Kerr black holes can acquire an additional global charge, called 'hair', which deviates from the Kerr metric \cite{dk}. Recently, the hairy black hole and later its rotating version were obtained using the gravitational decoupling approach (GD) \cite{gd0,gd}, which is specifically designed to describe deformations of known solutions of general relativity induced by additional sources\cite{gd1,gd2}. The rotating hairy black hole attracts amount of theoretical and observational investigations \cite{inv1,inv2,inv3,inv31,inv4,inv5}.

To verify this proposal in GW observations, the ringdown signal from black holes will be essential. This is because the ringdown waveforms arise from the gravitational perturbation of black holes, such as the final phase of black hole merger \cite{gw1,gw2,gw3,rd}, and they are the superposition of quasinormal modes which are directly related to the No Hair theorem. The discovery and accurate identification of the quasinormal modes could be the 'smoking gun' for testing this proposal as well as general relativity \cite{rd2}.

On the other hand, if the wavelength of gravitational perturbations is comparable to the size of the black hole horizon, they can efficiently extract rotational energy from rotating black holes through superradiance instabilities, which also generate GW and black hole shadow signals that could be detected by observations in the future \cite{sr,srs,srs1}. Because of its importance, superradiance has recently received a lot of attention from the scientific community, and physicists have explored a variety of aspects and scenarios.\cite{sr2,sr3,sr4,sr5,sr6,sr7,sr8,sr81,sr82,sr9,sr10,sr11,sr12,sr13,sr14,sr15,sr16,sr17,sr18,sr19,sr20,sr21}.

Studying the phenomenology of the gravitational perturbation field around rotating hairy black holes will therefore provide us with further insight into general relativity and gravity, and provide an interesting guide for future GW and black hole shadow observations. There is some work on this topic, but it is limited to scalar perturbations in the case of non-rotating hairy black holes \cite{inv3}. In this work, we will study the superradiance and quasinormal modes of the gravitational perturbation around the rotating hairy black hole \cite{gd}. We will show how the hairy parameters and the spin of the black hole affect the superradiance and the quasinormal modes.

This paper is organized as follows: In Sec.\ref{sec2}, we will introduce the rotating hairy black hole and derive the horizons using the approximation method. In Sec.\ref{sec3}, we obtain the master equations for the gravitational perturbation in this spacetime, and also drive the radial and angular equations respectively. Then, in Sec.\ref{sec4}, we will calculate the superradiance amplification factor and obtain the conditions for the superradiance. We will also explore the quasinormal modes in this spacetime in Sec.\ref{sec5}, after introducing the continued fraction method, we will present the numerical results on how the quasinormal modes will change as a function of the black hole spin, hairy parameters. In Sec.\ref{sec6}, we will come to a conclusion and discussion.

\section{rotating hairy black hole}\label{sec2}
In \cite{gd}, a rotating hairy black hole was derived using the GD approach which is precisely designed to find the deformation of the known solution of general relativity \cite{gd1,gd2}. In Boyer-Lindquist coordinates it is
\begin{equation}\label{metric}
\begin{aligned}
d s^{2}=&-\left[\frac{\Delta-a^{2} \sin ^{2} \theta}{\Sigma}\right] d t^{2}+\frac{\Sigma}{\Delta} d r^{2} +\Sigma d \theta^{2}\\
&-2 a \sin ^{2} \theta\left[1-\frac{\Delta-a^{2} \sin ^{2} \theta}{\Sigma}\right] d t d \phi \\
&+\sin ^{2} \theta\left[\Sigma+a^{2} \sin ^{2} \theta\left(2-\frac{\Delta-a^{2} \sin ^{2} \theta}{\Sigma}\right)\right] d \phi^{2}
\end{aligned}
\end{equation}
with $\Delta=r^{2}+a^{2}-2 M r+\alpha r^{2} e^{-r /\left(M-\frac{h_{0}}{2}\right)}$, and $\Sigma=r^{2}+a^{2} \cos ^{2} \theta$. $M$, $a$ denote the mass and spin of the black hole. $\alpha$ measures the deviation from standard Kerr black holes and is related to the primary hair $h_0$ via $h_0 = \alpha h$, where $h_0$ measures the increase of entropy caused by the hair and must satisfy the condition $h_0\le 2M\equiv h_K$ to ensure asymptotic flatness. The hair could originate from new fields or new gravitational sectors such as dark matter or dark energy as long as their effective energy-momentum tensor satisfies the strong energy condition outside the event horizon\cite{gd}. All these parameters were assumed to be positive. When $\alpha=0$, this spacetime reduces to the Kerr metric, which means the absence of surrounding matter.

The solutions of the equation
\begin{equation}\label{eh}
\Delta=r^{2}+a^{2}-2 M r+\alpha r^{2} e^{-r /\left(M-\frac{h_{0}}{2}\right)}=0
\end{equation}
will give us the horizons. One can find the numerical results of the horizon structure with different parameter values in \cite{inv2}. However, the exact analytical solutions are impossible for this $\Delta$ function.

Nevertheless, we can use the approximate method to solve (\ref{eh}) analytically as long as $\alpha e^{-r /\left(M-\frac{h_{0}}{2}\right)} \ll 1$, which means that the new field where the hair originate from is less dense, such that the deviation from the standard Kerr black hole is small, and it will also satisfy the condition for (\ref{eh}) to have two distinct real solutions, see \cite{inv2}.

For the Kerr black hole, $\Delta_{kerr}=r^{2}+a^{2}-2Mr=(r-r_+)(r-r_-)$, where $r_+$ and $r_-$ are the so-called event and inner horizon, respectively. They can be considered as the zero-order solution (in terms of $\alpha e^{-r /\left(M-\frac{h_{0}}{2}\right)}$) of equation (\ref{eh}). Because Eq. (\ref{eh}) can be written as
\begin{equation}\label{newapp}
r^{2}+a^{2}-2Mr=-\alpha r^{2} e^{-r /\left(M-\frac{h_{0}}{2}\right)}
\end{equation}
where the right side is much smaller than the left side if $ \alpha e^{-r /\left(M-\frac{h_{0}}{2}\right)} \ll 1$, so the right side is the small perturbation. Thus, if we bring the zero-order solutions $r_{\pm}$ into the right-hand side of (\ref{newapp}), we obtain high-order approximate solutions, they are
\begin{equation}\label{knapp}
\begin{array}{l}
\Delta_{kerr}+\alpha r_+^{2} e^{-r_+ /\left(M-\frac{h_{0}}{2}\right)}=(r-{r}_+^I)(r-\tilde{r}_-)\\
\Delta_{kerr}+\alpha r_-^{2} e^{-r_- /\left(M-\frac{h_{0}}{2}\right)}=(r-\tilde{r}_+)(r-{r}_-^I)
\end{array} 
\end{equation}
where ${r}_+^I$ and ${r}_-^I$ can be regarded as the first-order approximate solutions of (\ref{eh}), while $\tilde{r}_+$ and $\tilde{r}_-$ are the two additional roots, since we are solving two quadratic equations, and they are numerically less accurate compared to ${r}_+^I$ and ${r}_-^I$. The explicit forms for ${r}_+^I$ and ${r}_-^I$ are given by
\begin{align}
{r}_+^I=M +\sqrt{M^{2}-a^{2}-\alpha r_+^{2} e^{-r_+ /\left(M-\frac{h_{0}}{2}\right)}}\\
{r}_-^I=M -\sqrt{M^{2} -a^{2}-\alpha r_-^{2} e^{-r_- /\left(M-\frac{h_{0}}{2}\right)}}
\end{align}
For better accuracy, we could substitute ${r}_+^I$ and ${r}_-^I$ back to the right hand side of (\ref{newapp}) and repeat the above process to get more accurate second order solutions of (\ref{eh}),
\begin{align}\label{r++}
{r}_+^{II}=M +\sqrt{M^{2}-a^{2}-\alpha r_+^{I\,2} e^{-r_+^{I} /\left(M-\frac{h_{0}}{2}\right)}}\\
{r}_-^{II}=M -\sqrt{M^{2} -a^{2}-\alpha r_-^{I\,2} e^{-r_-^{I} /\left(M-\frac{h_{0}}{2}\right)}}\label{r--}
\end{align}
they could be considered as the event horizon and inner horizon of metric (\ref{metric}). One could repeat the approximation steps to obtain higher order solutions, but second order ${r}_+^{ II }$ and ${r}_-^{ II }$ are sufficient in this work, see Appendix.\ref{ap1}. This approximation method has been used previously to study rotating regular black holes in \cite{zhen}. In the following, for simplicity, we will define $\hat r_+ \equiv {r}_+^{ II }$ and $\hat r_- \equiv {r}_-^{ II }$. Then the $\Delta$ function could be approximately written as
\begin{equation}\label{eh2}
\Delta\approx(r-\hat r_+)(r-\hat r_-)
\end{equation}
In the following sections, whenever we write $\Delta$ for metric (\ref{metric}), we refer to the above function (\ref{eh2}).

\section{decoupled perturbation equations}\label{sec3}
The spacetime symmetries and the asymptotic behavior of this rotating hairy black hole (\ref{metric}) are the same to Kerr black hole\cite{nre2}, so we can decompose the perturbation field $\Phi$ with the ansatz
\begin{equation}\label{ansa}
\Phi\left(x^{\mu}\right)=e^{-i \omega t} e^{i m \phi}S_{sl m}(\theta) R_{slm}(r)
\end{equation}
where $\omega$ is the complex-valued frequency, $l$, $m$ are the quantum numbers.

In a breakthrough work\cite{tes}, it was shown that the linearized gravitational perturbations of Kerr geometry can be described by a master equation: the Teukolsky equation. However, the Teukolsky equation does not apply to the metric (\ref{metric}) because it is not a solution of Einstein's field equation. Fortunately, using the approximation for the $\Delta$ function in Sec.\ref{sec2}, we can see from Eq.(\ref{knapp}) that it effectively corresponds to the $\Delta$ function of a Kerr-Newman black hole, with the effective charge $Q_{effect}\approx \sqrt{ \alpha \hat r_+^2 e^{-\hat r_+ /\left(M-\frac{h_{0}}{2}\right)}}$ or $\sqrt{\alpha \hat r_-^2 e^{-\hat r_- /\left(M-\frac{h_{0}}{2}\right)}}$ (we used the results of the second order approximation). For a Kerr-Newman black hole, it has been shown that the gravitational perturbation can be described by the Dudley-Finley equation \cite{df}, which provides a good approximation to the dynamics of the perturbation field when the electric charge $Q \le M/2$ \cite{kk,kk1}. The condition is also satisfied in our situation, since we assume $Q_{effect}\ll M$ by the definition of the approximation method in Sec.\ref{sec2}.

The final $\Delta$ function (\ref{eh2}) we considered could be seen as a Kerr-Newman black hole with effective mass $M_{effect}=(\hat r_++\hat r_-)/2$ and charge $Q_{effect}=\sqrt{\hat r_+\hat r_--a^2}$. It could be proved that $Q_{effect}\ll M_{effect}$. Therefore, we can approximately obtain a pair of differential equations from the Dudley-Finley equation \cite{kk,kk1}, and in fact these two equations have the same form as the Teukolsky equations, with the only difference that we replace $\Delta_{kerr}$ by $\Delta$ function (\ref{eh2}) compared to the Kerr case. The radial equation has the following form
\begin{equation}\label{radial}
\begin{aligned}
&\Delta^{-s}\frac{d}{d r}\left(\Delta^{s+1} \frac{d R_{slm}}{d r}\right)\\ &+\left(\frac{K^2-is K\Delta'}{\Delta}+4is\omega r -\lambda\right) R_{slm}(r)=0
\end{aligned}
\end{equation}
where $s$ is the spin weight of the
gravitational field, $K=(r^2+a^2)\omega-am$ and $\lambda=\Lambda_{l m}+a^2\omega^2-2am\omega$, $\Lambda_{l m}$ is the separation constant, they are the eigenvalues with respect to the following angular equation,
\begin{align}\label{angular}
&\frac{1}{\sin \theta} \frac{d}{d \theta}\left(\sin \theta \frac{d S_{sl m}}{d \theta}\right)+\left(a^{2} \omega^{2}\cos ^{2} \theta-\frac{m^{2}}{\sin ^{2} \theta}-2s a\omega\cos  \theta\right. \nonumber\\
&\left.-\frac{2s m\cos  \theta}{\sin^2\theta}-s^2\cot^2\theta+s+\Lambda_{l m}\right) S_{sl m}(\theta)=0
\end{align}
The solutions of the angular equation are $S_{l m}(\theta)$, which are called spheroidal harmonics\cite{asf}. In the nonrotating or spherically symmetric limit case, the spheroidal harmonics reduce to the well-known spherical harmonics $Y_{l m}$ and also $\Lambda_{l m}\approx l(l+1)- s(s+1)$.

We define $u(r) \equiv \sqrt{r^{2}+a^{2}} R_{slm}(r)$ and switch to the tortoise coordinate via $d r_{*}=\frac{r^{2}+a^{2}}{\Delta} d r$, after some algebra, the radial function (\ref{radial}) takes the following Schr\"odinger-like form
\begin{equation}\label{u}
\frac{d^{2} u\left(r_*\right)}{d r_*^ 2}+\mathcal{V}(r) u\left(r*\right)=0
\end{equation}
with the effective potential $\mathcal{V}(r)$ given by
\begin{widetext}
\begin{align}\label{po}
\mathcal{V}(r)=&\left(\omega-\frac{a m}{a^{2}+r^{2}}\right)^{2}-\frac{i s\omega \Delta'}{a^{2}+r^{2}}+\frac{i s a m \Delta'}{(a^{2}+r^{2})^2}+\frac{\Delta }{({a^{2}+r^{2}})^2}\left (4 i s\omega r-\lambda \right )- \frac{d G}{d r_*}-G^2
\end{align}
\end{widetext}
where $G= r\Delta/(r^2+a^2)^2+s\Delta'/2(r^2+a^2) $. The last two terms represent the effect of introducing the tortoise coordinate $dr_{*}$. The other terms result from the potential of Eq.(\ref{radial}) divided by $(r^2 + a^2)^2$.

\section{superradiance and amplification factor}\label{sec4}
Within a certain black hole parameter space, the incident gravitational waves can be amplified if they are scattered by a rotating black hole. In this way, the gravitational waves can efficiently extract the rotational energy of the black hole, i.e., the gravitational waves of the black hole are superradiant.
We focus on the regime in which the backreaction of the superradiant waves on the geometry is negligible, i.e., the background spacetime metric is held fixed. This can be easily satisfied since we are working in the linear perturbation framework, and furthermore the superradiant waves are typically distributed over a very large area, which implies very low density and consequently small backreaction effects. The work \cite{extr} which investigate the backreaction of bosonic clouds on the Kerr geometry could be a good reference.

In this section we will examine the conditions for the occurrence of superradiance. We consider the following boundary conditions of Eq.(\ref{u}),
\begin{equation}\label{sb}
\begin{aligned}
u_{h}(r_*) &=\mathcal{T}_{s}\Delta^{-s/2} \exp \left(-i k_{h} r_{*}\right), r_{*} \longrightarrow -\infty(r \rightarrow \hat{r}_{+}) \\
u_{\infty}(r_*) &=\mathcal{I}_{s} r^{s} \exp \left(-i \omega r_{*}\right)\\
&+\mathcal{R}_{s}r^{-s}\exp \left(i  \omega r_{*}\right),  r_{*} \longrightarrow \infty(r \rightarrow \infty)
\end{aligned}
\end{equation}
where $k_{h}=\omega-m\Omega_{h}$, $\Omega_h=a/(\hat{r}_+^2+a^2)$.
The above boundary conditions state that there is an incoming wave from spatial infinity with amplitude $\mathcal{I}_{s}$ that is scattered at the event horizon such that it produces reflected waves with amplitude $\mathcal{R}_{s}$ and transferred waves with amplitude $\mathcal{T}_{s}$ respectively. The subscription $s$ means that all amplitudes are $s$ dependent.

To quantify how much the incident waves were amplified due to superradiance, one usually refers to the amplification factor $Z_{slm}$, which is given by \cite{sr,s-s}
\begin{equation}\label{aff}
Z_{sl m}=\left|\frac{\mathcal{R}_{s}\mathcal{R}_{-s}}{\mathcal{I}_{s}\mathcal{I}_{-s}}\right|-1
\end{equation}
where $\mathcal{R}_{-s}$ and $\mathcal{I}_{-s}$ means the substitution of $s$ by $-s$ in the solution of Eq.(\ref{u}) with the asymptotic behavior of $u_{\infty}(r_*)$ given by
\begin{equation}
\begin{aligned}
u_{\infty}(r_*) &=\mathcal{I}_{-s} r^{-s}\exp \left(-i \omega r_{*}\right)\\
&+\mathcal{R}_{-s}r^{s}\exp \left(i  \omega r_{*}\right),  r_{*} \longrightarrow \infty(r \rightarrow \infty)
\end{aligned}
\end{equation}
The derivation of $Z_{slm}$ in this rotating hairy black hole (\ref{metric}) spacetime is about the same as in Kerr\cite{sr}, with the difference that $r_+$ is replaced by $\hat{r}_ +$. However, this changes the geometry of the region near the horizon, which also affects the amplification factor. Just to be self-content, we shall briefly discuss the calculation process of amplification factor $Z_{slm}$.

We will consider the low frequency range $ \omega M \ll 1$  which also implies $ a \omega \ll 1$. These approximations allow us to use the matching-asymptotic techniques \cite{match,match1,match2} as follows.

We divide the space into two overlapping regions, which we call the near region $\omega (r- r_+) \ll 1$ and the far region $ r-r_+\gg M$. Then
We can solve the radial Eq. (\ref{radial}) in these two regions and match the solutions in the overlapping region. In this way, we obtain the analytical solutions for the amplitudes, which we can use to calculate the amplification factor. The specific steps are as follows.

Let's rewrite (\ref{radial}) as
\begin{widetext}
\begin{align}\label{xradial}
x^2(1+x)^{2} \frac{d^{2} R_{slm}}{d x^{2}}&+ x(x+1)(2 x+1) \frac{d R_{slm}}{d x}+(k^{2} x^{4}+2i s k x^3 -\lambda x(x+1)-isQ(2x+1)+Q^{2}) R_{slm}=0
\end{align}
\end{widetext}
where we have defined new variables 
\begin{align}
x&=\frac{r-\hat r_+}{\hat r_+-\hat r_-} \\
k&=\omega(\hat r_+-\hat r_-)\\
Q&=\frac{\hat{r}_{+}^{2}+a^{2}}{\hat{r}_{+}-\hat{r}_{-}}(m \Omega_h-\omega)
\end{align}

In the near region, the quantity $kx \ll 1 $, so Eq.(\ref{xradial}) could be approximately written as
\begin{align}
&x^{2}(x+1)^{2} \frac{d^{2}R_{slm}}{d x^{2}}+x(x+1)(2 x+1) \frac{d R_{slm}}{d x}\nonumber\\
&+\left(Q^{2}-i s Q(2x+1)-l(l+1) x(x+1)\right) R_{slm}=0
\end{align}
the general solutions to the above equation as well as satisfying the boundary conditions (\ref{sb}) are given by the hypergeometric functions
\begin{align}
R_{slm}=A_{1} (\frac{x+1}{x})^{-s+i Q} F(\xi, \beta, \gamma,-x)
\end{align}
where 
\begin{align}
\gamma&=1-s-2i Q\\
\beta&=l-s+1\\
\xi&=-l-s
\end{align}

the large $x$ behavior of above solution is
\begin{align}\label{x}
R_{slm} &\sim A_{1}x^{l-s} \frac{\Gamma(\gamma) \Gamma(\beta-\xi)}{\Gamma(\gamma-\xi) \Gamma(\beta)}\nonumber\\
&+A_{1}x^{-l-1-s} \frac{\Gamma(\gamma) \Gamma(\xi-\beta)}{\Gamma(\xi) \Gamma(\gamma-\beta)}
\end{align}

In the far region, equivalently $x\rightarrow \infty$, Eq.(\ref{xradial}) approximately give us
\begin{equation}
\frac{d^{2}{R_{slm}}}{d x^{2}}+\frac{2}{x} \frac{{d}{R_{slm}}}{{d} x}+\left(k^{2}+\frac{2i s k}{x}-\frac{l(l+1)}{x^{2}}\right) {R_{slm}}=0
\end{equation}
The solution of this equation can be written in terms of the confluent hypergeometric function
\begin{align}\label{kk}
{R}_{slm}&=\exp (-i k x)C_{1} x^{l-s} U(l-s+1,2 l+2,2 i k x)\nonumber\\
&+\exp (-i k x)C_{2} x^{-l-1-s} U(-l-s,-2 l, 2 i k x)
\end{align}
Applying the condition $kx \ll 1$ and Expanding it with respect to $kx$, we obtain
\begin{equation}\label{k}
{R}_{slm}\sim C_{1} x^{l-s}+C_{2} x^{-l-1-s}
\end{equation}
Now, we can match the solutions (\ref{x}) and (\ref{k}), and it will give us
\begin{align}
C_{1}=A_{1} \frac{\Gamma(1-s-2 i Q) \Gamma(2 l+1)}{\Gamma(l+1-s) \Gamma(l+1-2 i Q)} \nonumber\\
C_{2}=A_{1} \frac{\Gamma(1-s-2 i Q) \Gamma(-1-2 l)}{\Gamma(-l-2 i Q) \Gamma(-l-s)} \nonumber
\end{align}

From (\ref{sb}), we can know that, when $r\rightarrow \infty$, the solution of (\ref{radial}) will take form as
\begin{equation}\label{rin}
{R}_{slm}\sim \frac{u_{\infty}(r_*)}{r} \sim \mathcal{I}_{s} \frac{\exp \left(-i \omega r_{*}\right)}{r}+\mathcal{R}_{s}\frac{\exp \left(i \omega r_{*}\right)}{r^{2s+1}}
\end{equation}
Expanding (\ref{kk}) at infinity and matching to (\ref{rin}), we obtain the analytical expression for $\mathcal{I}_{s}$ and $\mathcal{R}_{s}$
\begin{align}
\mathcal{I}_{s} &=C_{1} \frac{(-2 i)^{s-l-1} k^{s-l} \Gamma(2 l+2)}{\omega \Gamma(l+s+1)}\nonumber \\
&+C_{2} \frac{(-2 i)^{l+s} k^{l+1+s} \Gamma(-2 l)}{\omega \Gamma(-l+s)}\nonumber \\
\mathcal{R}_{s} &=C_{1} \frac{(2 i)^{-l-1-s} k^{s-l} \Gamma(2 l+2)}{\omega^{2s+1}\Gamma(l+1-s)}\nonumber \\
&+C_{2} \frac{(2 i)^{l-s} k^{l+1+s} \Gamma(-2 l)}{\omega^{2s+1}\Gamma(-l-s)}.
\end{align}

\begin{figure}[htbp]
\centering
\includegraphics[scale=0.6]{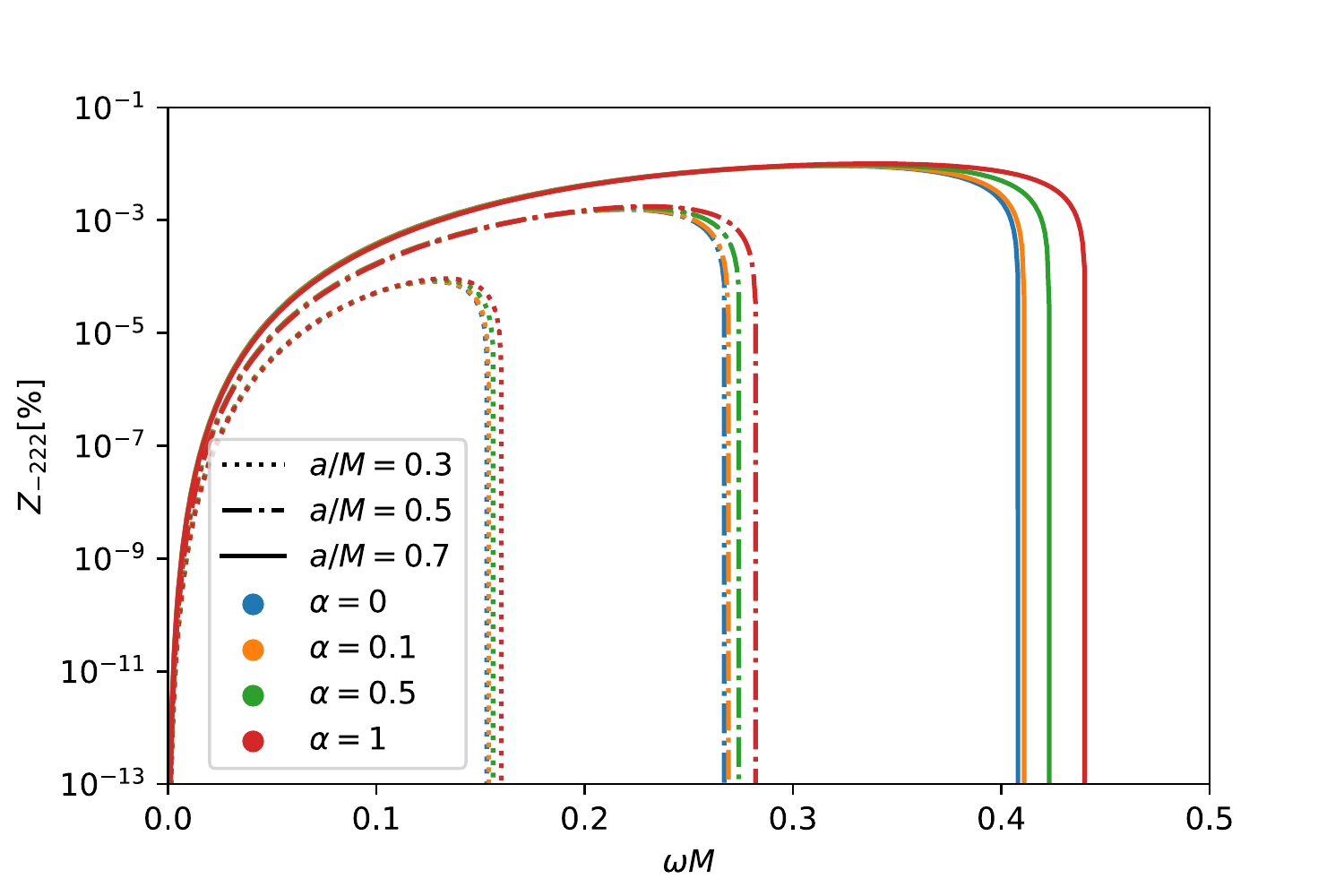}
\includegraphics[scale=0.6]{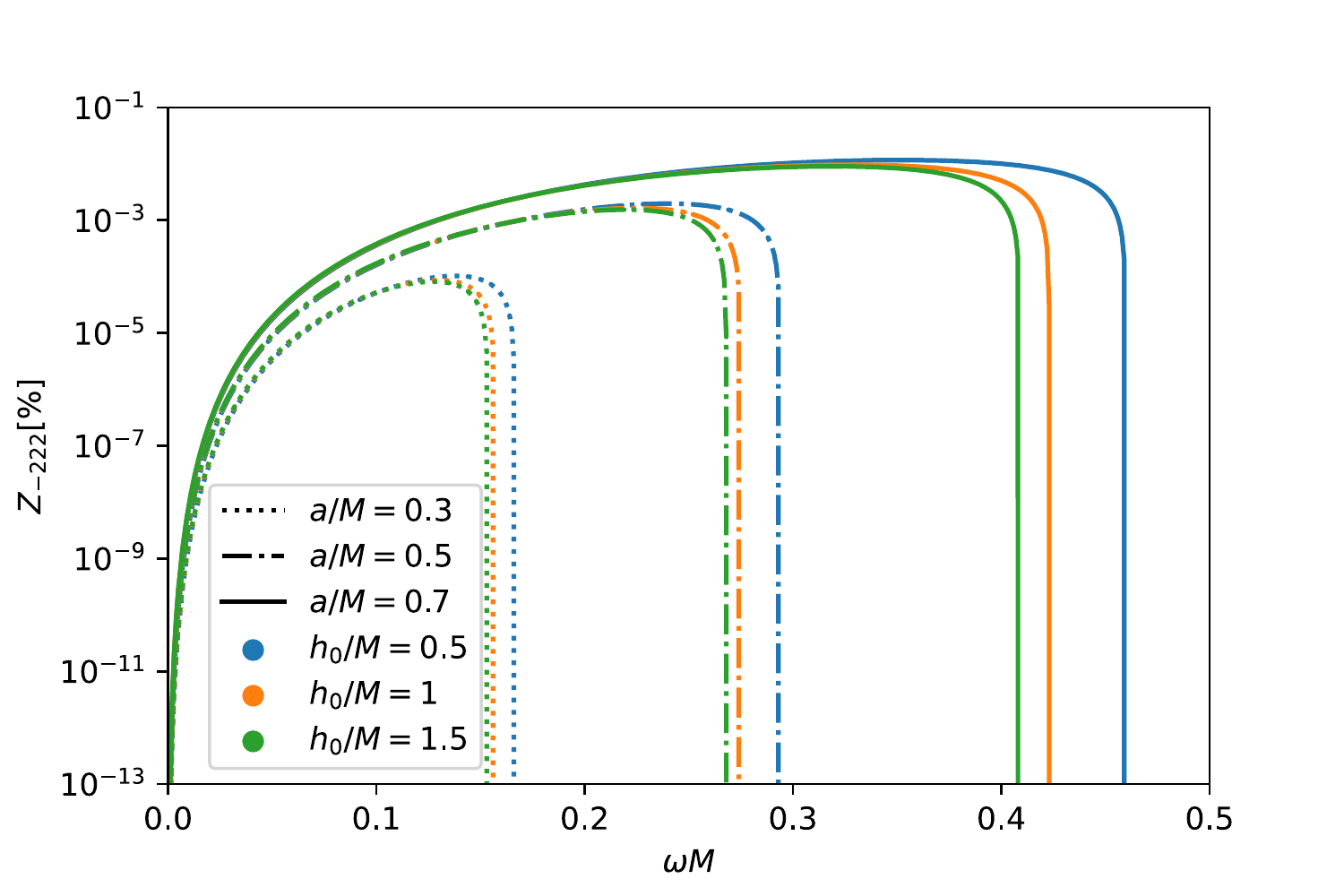}
\caption{The amplification factor $Z_{-222}$ for $s=-2,l=m=2$, with three black hole spin $a=0.3M$, $a=0.5M$ and $a=0.7M$. The upper plot aims to show the effects of $\alpha$ to the amplification by setting $h_0=M$ with $\alpha\ne 0$ (please note that $\alpha=0$ also means $h_0=0$). The lower plot is to show the effects of $h_0$ to the amplification by setting $\alpha=0.5$.}
\label{af}
\end{figure}

After some calculations and algebra, we finally find that the amplification factor (\ref{aff}) takes the following explicit form
\begin{equation}\label{ll}
Z_{slm}=4 Q k^{2 l+1}\frac{((l-s) !(l+s)!)^{2}}{((2 l) !)^{2}((2 l+1) ! !)^{2}} \\
\times \prod_{n=1}^l\left(1+\frac{4 Q^{2}}{n^{2}}\right)
\end{equation}
The above formula is applicable to any spin $a \le M $ as long as $ \omega M \ll 1$. In Fig.\ref{af}, $Z_{-222}$ is shown for different values of the parameter $\alpha,h_0$ and the spin of the black hole. We can clearly see that the amplification starts at $\omega M > 0$, and stops abruptly when it comes close to the threshold frequencies $m\Omega_h$. Increasing the spin of the black hole leads to an increase in the amplification, and the parameters $\alpha$ and $h_0$ only affect the amplification when the frequencies reach the threshold $m\Omega_h$. Larger $\alpha$ and smaller $h_0$ cause a larger threshold frequency.

The formula (\ref{ll}) also gives us the conditions of superradiance. If we want the amplitude of the reflected waves to be larger than that of the incident waves, i.e. $Z_{slm} > 0$ which means $Q > 0$, the following frequency criteria must be fulfilled
\begin{equation}\label{cds}
0<\omega<m\Omega_{h}
\end{equation}
where $0<\omega$ is required by the boundary conditions. The modes that satisfy the above condition are also called superradiance modes.

One could also study the amplification of scalar and vector perturbation fields using the same formula (\ref{ll}) by simply substituting $s=0$ and $s=\pm1$. However, we will focus only on amplification of gravitational perturbations because it is more interesting and could potentially provide phenomenological intuitions for gravitational wave observations. In the next section, we will also study the quasinormal modes of gravitational perturbations, which are closely related to the ringdown signal of gravitational waves.

\section{quasinormal Modes}\label{sec5}
Quasinormal modes are the solutions of the equation (\ref{u}) satisfying the following boundary conditions,
\begin{align}\label{b2}
u_{h}(r_*) &\propto  \Delta^{-s/2}\exp \left(-i k_{h} r_{*}\right),\quad r_{*} \longrightarrow -\infty(r \rightarrow \hat{r}_+) \nonumber\\
u_{\infty}(r_*) &\propto  r^{-s}\exp \left(i \omega r_{*}\right), \quad r_{*} \longrightarrow \infty(r \rightarrow \infty)
\end{align}
These boundary conditions mean that there are only incoming waves at the event horizon, while there are pure outgoing waves at spatial infinity. These conditions lead to discrete eigenvalue frequencies. Quasinormal modes are also called "fingerprints" of black holes. This is because they are completely determined by the hair parameters of the black hole, such as mass and spin.

There are a number of different approaches to calculating quasinormal modes, see reviews in \cite{qnm,qnm1,qnm2}. In this work we will use the popular continued fraction method, which has been used in many outstanding papers, including recent ones\cite{cfm,cfm1,cfm2,zhen}.

\subsection{Continued fraction method}
According to the boundary conditions (\ref{b2}) and the poles of the radial Eq. (\ref{radial}) we can formulate a series solution for $R_{slm}(r)$ as
\begin{equation}\label{se1}
R_{slm}(r)=\left(r-\hat r_{+}\right)^{-i \sigma_{+}}\left(r-\hat r_{-}\right)^{i \sigma_{-}}y(r-\hat r_-)
\end{equation}
where $-i\sigma_{+}$ and $i\sigma_{-}$ are the indices of $R_{sl m}(r)$ corresponding to the singular points $r=\hat r_+$ and $r=\hat r_-$ respectively, they have the form
\begin{align}
-i\sigma_{+}=-s-i\frac{\omega \hat r_+ - am}{b}\\
i\sigma_{-}=-s+i\frac{\omega \hat r_+ - am}{b}
\end{align}
where $b=\hat r_+-\hat r_-$. For simplicity, we define $x=r-\hat r_-$ and thus $\Delta=x(x-b)$. Then we can rewrite (\ref{se1}) and (\ref{radial}) into the expression of $x$ and substitute the series solution (\ref{se1}) into (\ref{radial}). Next, we compute the derivatives in (\ref{radial}). Combining the terms based on $y(x)$, $\frac{d y}{d x}$, and $\frac{d^{2} y}{d x^{2}}$, respectively, we obtain the equation for $y(x)$,
\begin{align}\label{yy}
x\left(x-b\right) \frac{d^{2} y}{d x^{2}}&+\left(B_{1}+B_{2} x\right) \frac{d y}{d x}+\left(\omega^{2} x\left(x-b\right) \right. \nonumber\\
&\left.-2 \eta \omega\left(x-b\right)+B_{3}\right) y=0
\end{align}
where $B_1$, $B_2$, $B_3$ and $\eta$ are given by
\begin{align}
 B_1 &= (-s-1-2i\sigma_{-})b\nonumber\\
 B_2 &= 2(i\sigma_{-}-i\sigma_{+}+s+1)\nonumber\\
 B_3 &= 2\omega^2\hat r_+^2+\omega^2(\hat r_++\hat r_-)^2+a^2\omega^2-\Lambda_{l m}\nonumber\\
 &+i(2s+1)(\sigma_{-}-\sigma_{+})-(\sigma_{-}-\sigma_{+})^2-is\omega(b+2-4\hat r_+)\nonumber\\
 \eta &= -\omega(\hat r_+ + \hat r_-)-is\nonumber
\end{align}
The function $y(x)$ can be extended to a power series in the following form
\begin{equation}\label{se2}
y(x)=e^{i \omega x} x^{-(1 / 2) B_{2}-i \eta} \sum_{n=0}^{\infty} d_{n}\left(\frac{x-b}{x}\right)^{n} 
\end{equation}
Substituting this new power series solution of $y(x)$ into Eq.(\ref{yy}). We obtain the coefficients $d_n$ satisfying a three term recurrence relation as follows
\begin{align}\label{cf1}
\alpha_{0} d_{1} +\beta_{0} d_{0}& =0 \\
\alpha_{n} d_{n+1}+\beta_{n}d_{n}+\gamma_{n}d_{n-1}& =0, \quad n=1,2,3,... \ldots\label{cf2}
\end{align}
where the coefficients are
\begin{align}
\begin{array}{l}
\alpha_{n}=n^{2}+\left(c_{0}+1\right) n+c_{0} \\
\beta_{n}=-2 n^{2}+\left(c_{1}+2\right) n+c_{3} \\
\gamma_{n}=n^{2}+\left(c_{2}-3\right) n+c_{4}-c_{2}+2
\end{array}
\end{align}
and the intermediate constant $c_n$ are defined as
\begin{equation}
\begin{array}{l}
c_{0}=B_{2}+B_{1} / b\\ c_{1}=-2\left(c_{0}+1+i\left(\eta-\omega b\right)\right) \\
c_{2}=c_{0}+2(1+i \eta) \\
c_{3}=-c_{4}-\frac{1}{2} B_{2}\left(\frac{1}{2} B_{2}-1\right)+\eta(i-\eta)+i \omega b c_{0}+B_{3} \\
c_{4}=\left(\frac{1}{2} B_{2}+i \eta\right)\left(\frac{1}{2} B_{2}+i \eta+1+B_{1} / b\right) \\
\end{array}
\end{equation}
By combining Eq.(\ref{cf1}) and (\ref{cf2}), we can get
\begin{equation}\label{cf}
0=\beta_{0}-\frac{\alpha_{0} \gamma_{1}}{\beta_{1}-} \frac{\alpha_{1} \gamma_{2}}{\beta_{2}-} \frac{\alpha_{2} \gamma_{3}}{\beta_{3}-} \ldots
\end{equation}
If the power series (\ref{se1}) and (\ref{se2}) converge and satisfy the boundary condition $r=\infty$ (\ref{b2}), then for a given $a$, $M$, $\alpha$, $h_0$, and $\Lambda_{l m}$ the frequency $\omega$ must be a root of the continued fraction Eq.(\ref{cf}) or any of its inversions. The solutions of Eq.(\ref{cf}) give us the quasinormal modes.

\subsection{Numerical results}
For simplicity and in agreement with the literature, we set $M = 1$ as the unit in the rest of this paper. Then the radial distance $r$, the hair parameter $h_0$ and the spin of the black hole $a$ are measured in the unit $M$, while the frequency $\omega$ is given in the unit $M^{-1}$.

Our numerical procedures are as follows: first, we compute the angular eigenvalues using the Leaver method\cite{lea}, by fixing the values for ($\alpha$, $h_0$, $l$, $m$, $a$). Then the continued fraction Eq.(\ref{cf}) is only a function of the quasinormal modes $\omega$. It is more convenient and necessary to truncate the continuing fraction to a certain order $N$, so we use the technique developed by Nollert \cite{nl} to fix the value of $N$. Finally, the root-finding algorithm (built-in functions in \textit{Wolfram Mathematica}) is used to find the roots of Eq. (\ref{cf}). Previous calculations of quasinormal modes in the Kerr background and also the scalar field perturbation scenario \cite{valid,valid1,valid2} are used to validate and verify our numerical methods. The errors of the quasinormal modes caused by using the approximation (\ref{r++}) and (\ref{r--}) are smaller than $10^{-2}$, see Appendix.\ref{ap1}, they become extremely accurate in the slow rotation region.

In Table.\ref{tab1} we show some of the quasinormal modes for the fundamental mode $(s=-2,l=m=2)$ with different black hole spins $a$, parameter $\alpha$ by setting $h_0 =1$. The columns with $\alpha=0$ correspond to the quasinormal modes of the Kerr black hole, which are in excellent six decimals agreements with the results previously obtained by \cite{valid}. In Table.\ref{tab2} we show some of the $(s=-2,l=m=2)$ quasinormal frequencies for the fundamental mode with different black hole spins and parameter $h_0$ by setting $\alpha =0.5$. In Table.\ref{tab3} we show few overtones of $(s=-2,l=m=2)$ quasinormal frequencies with different black hole spins $a$, by setting $h_0 =1$ and $\alpha=0.5$.

From Table.\ref{tab1}, Table.\ref{tab2}, and Table.\ref{tab3}, we notice that the real part of the quasinormal frequencies increases with the spin of the black hole, while the imaginary part decreases with the spin, regardless of the parameters and overtones are. This can be caused by the choice of a certain quantum number $l=m=2$.

\begin{figure}[htbp]
   \includegraphics[scale=0.6]{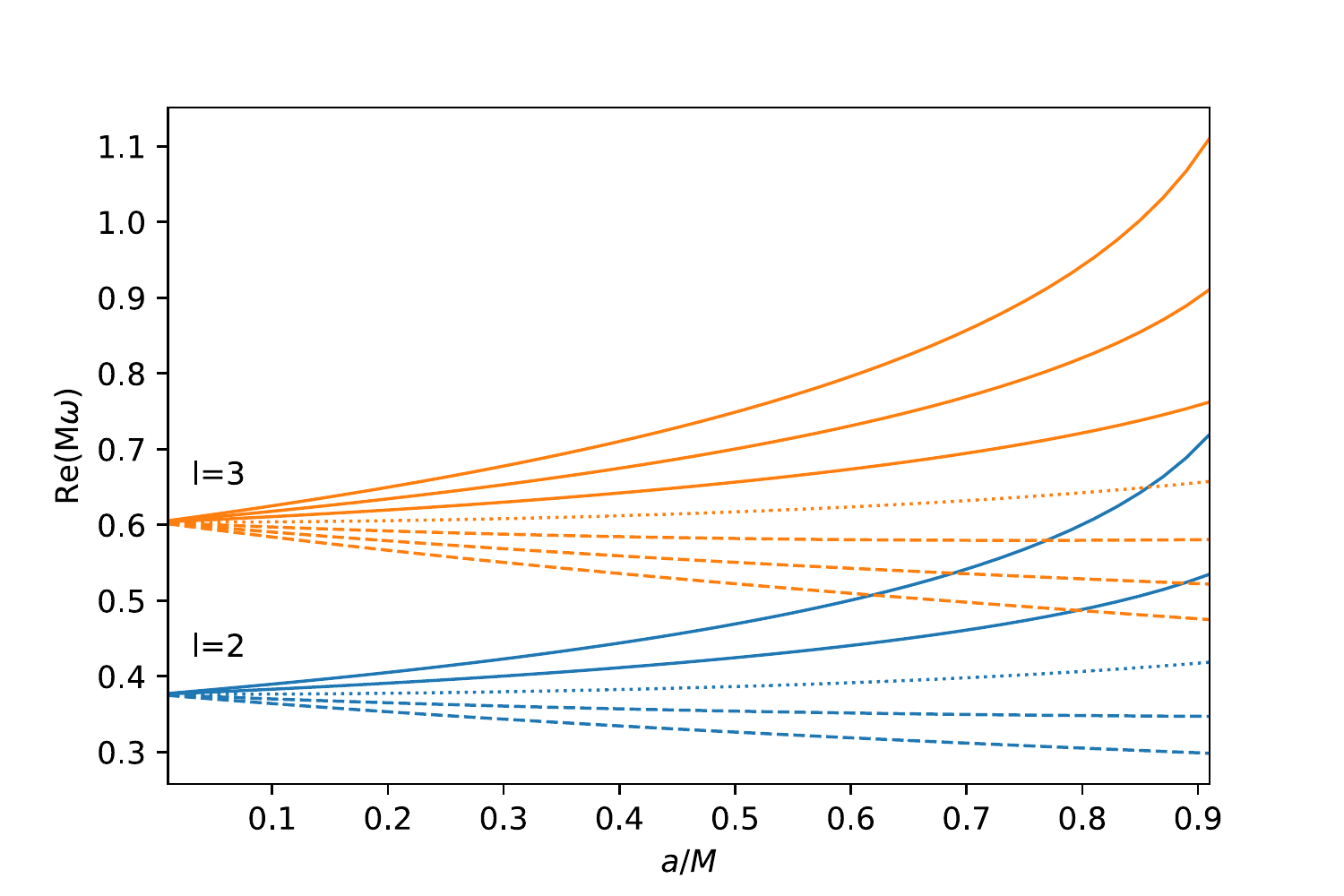}
   \includegraphics[scale=0.6]{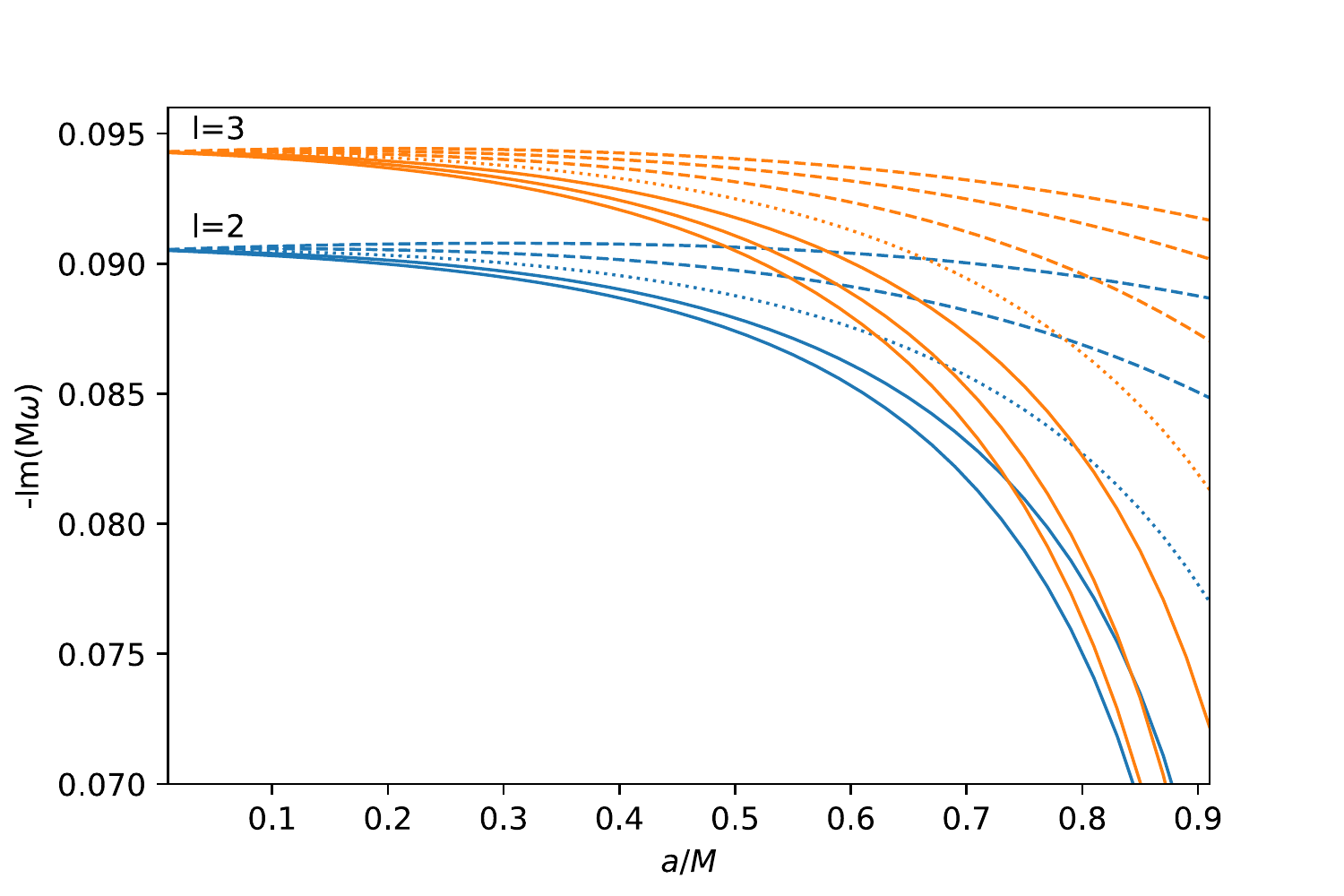}
  \caption{The $s=-2$ fundamental quasinormal modes with $l =$2, 3, $\alpha=0.5$, $h_0=1$ and different values of m. The upper panel is the real part. From top to bottom, solid lines correspond to $m = l,...,1$, dotted line $m = 0$, and dashed lines correspond to $m = -1,..., -l$. The low panel is the minus imaginary part, From top to bottom, solid lines correspond to $m = 1,...,l$, dotted line $m = 0$, and dashed lines correspond to $m = -l,...,-1$. }
\label{lm}
\end{figure}

\begin{figure}[htbp]
   \includegraphics[scale=0.6]{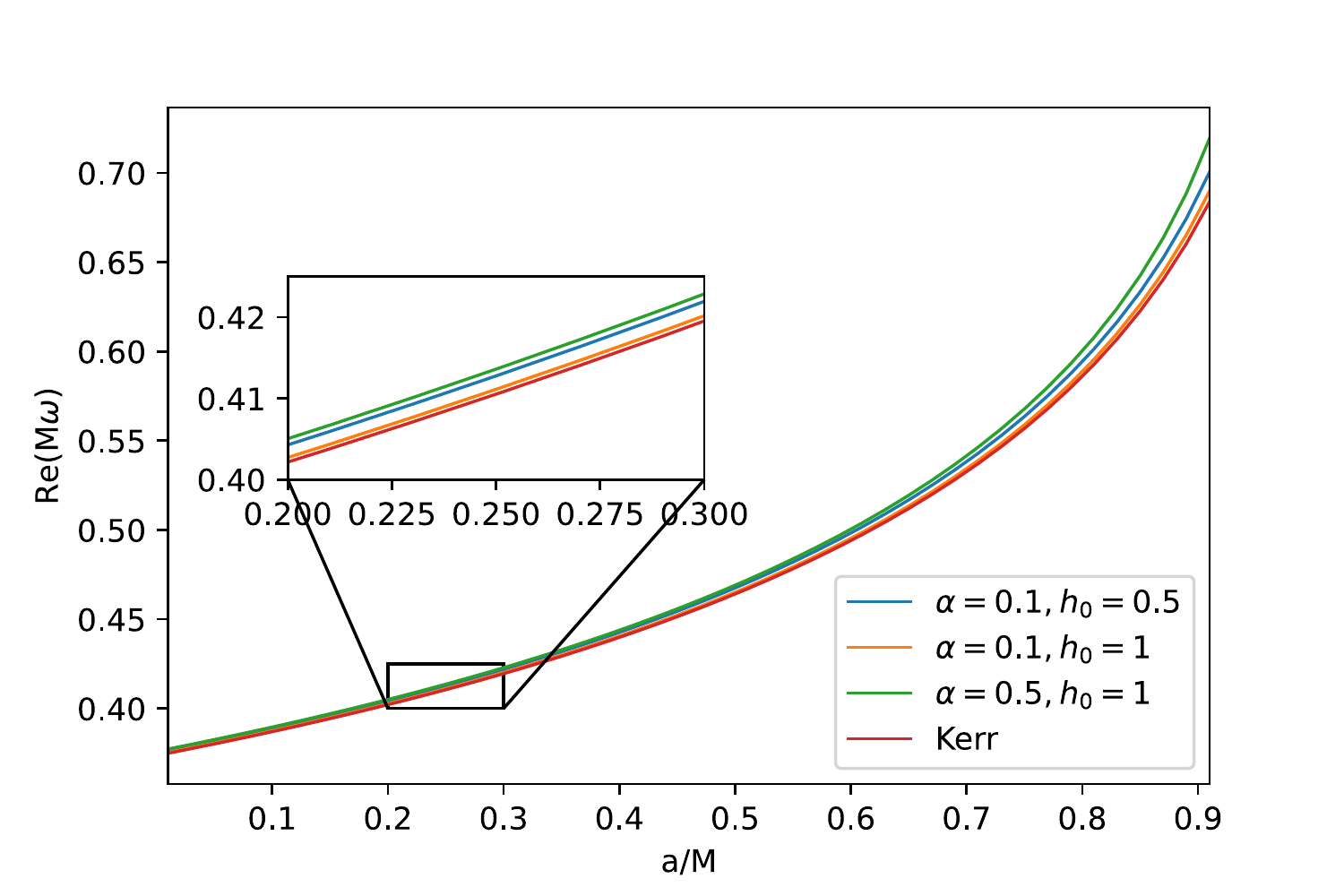}
   \includegraphics[scale=0.6]{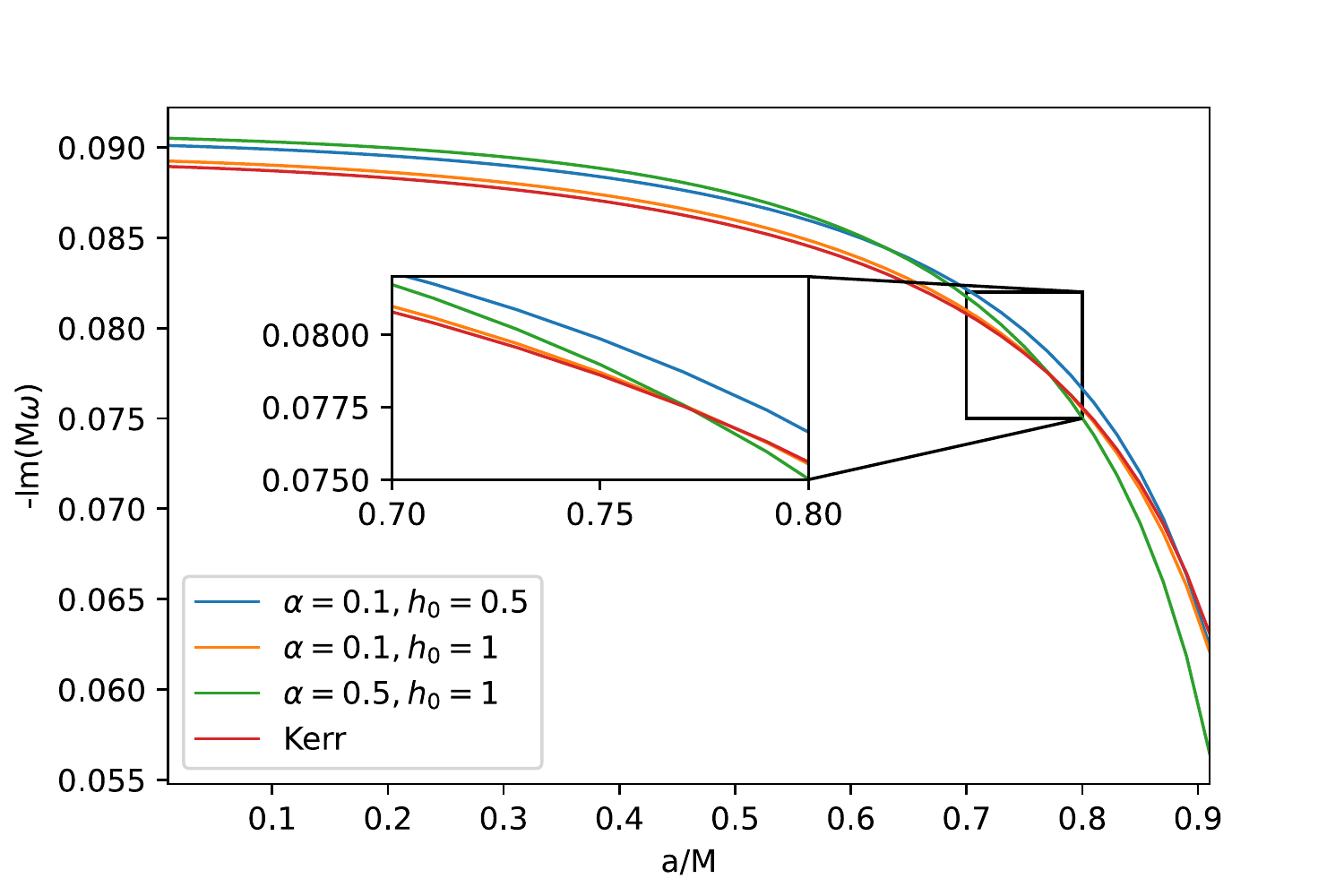}
  \caption{The $s=-2$ fundamental quasinormal modes with $l=m=2$ as a function of spin $a$, with different  combinations of $\alpha$, $h_0$. The upper panel is the real part while the low panel is the minus imaginary part }
\label{reim-a}
\end{figure}

\begin{figure}[htbp]
   \includegraphics[scale=0.6]{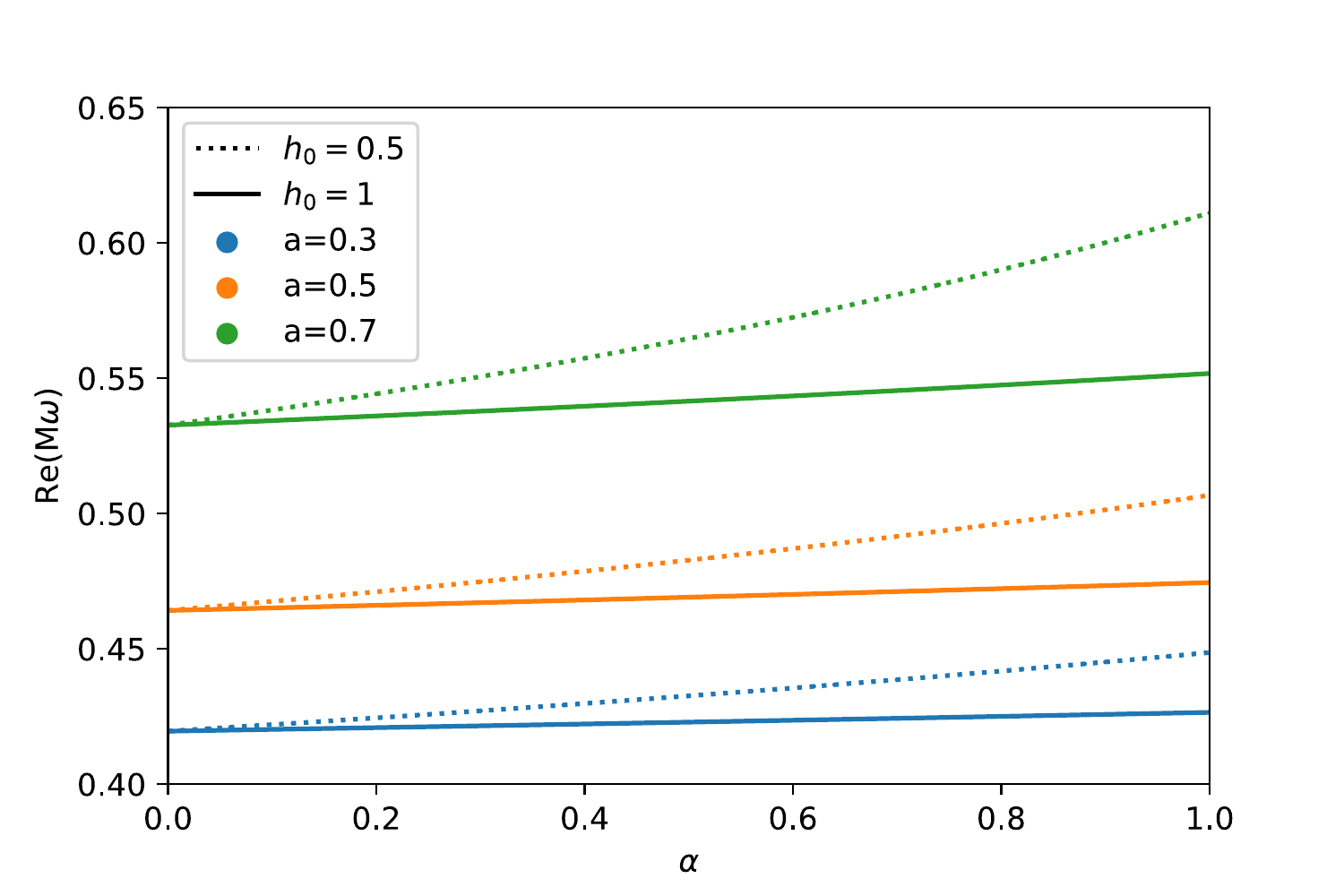}
   \includegraphics[scale=0.6]{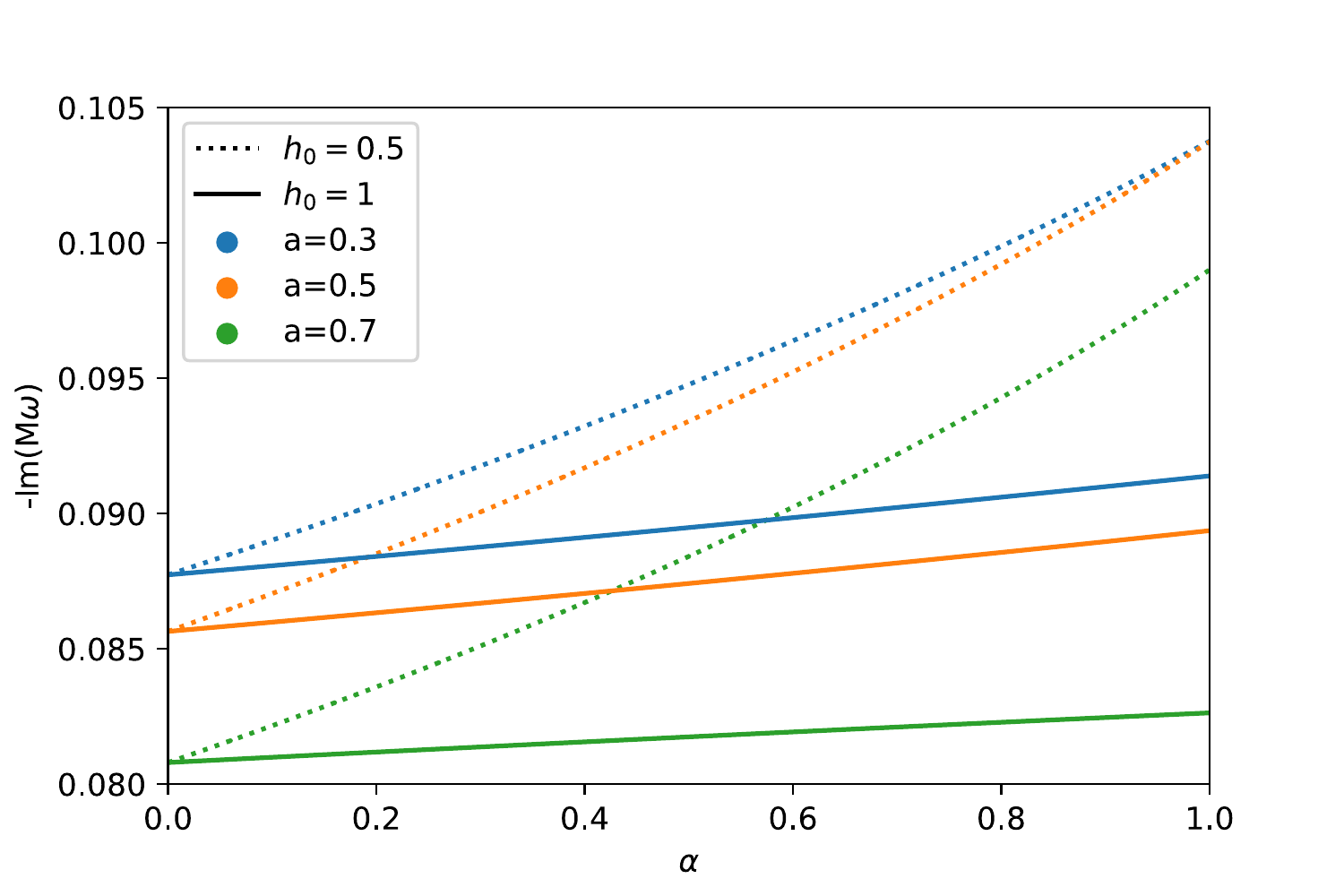}
  \caption{The real and imaginary part of $s=-2, l=m=2$ fundamental quasinormal modes as a function of parameter $\alpha\in[0,1]$, with different spin $a$ and $h_0$ (please note when $a\le0.7$, the errors are also below $10^{-2}$ even with $\alpha=1, h_0=0.5$, see Fig.\ref{err})}
\label{reim-alpha}
\end{figure}

\begin{figure}[htbp]
   \includegraphics[scale=0.6]{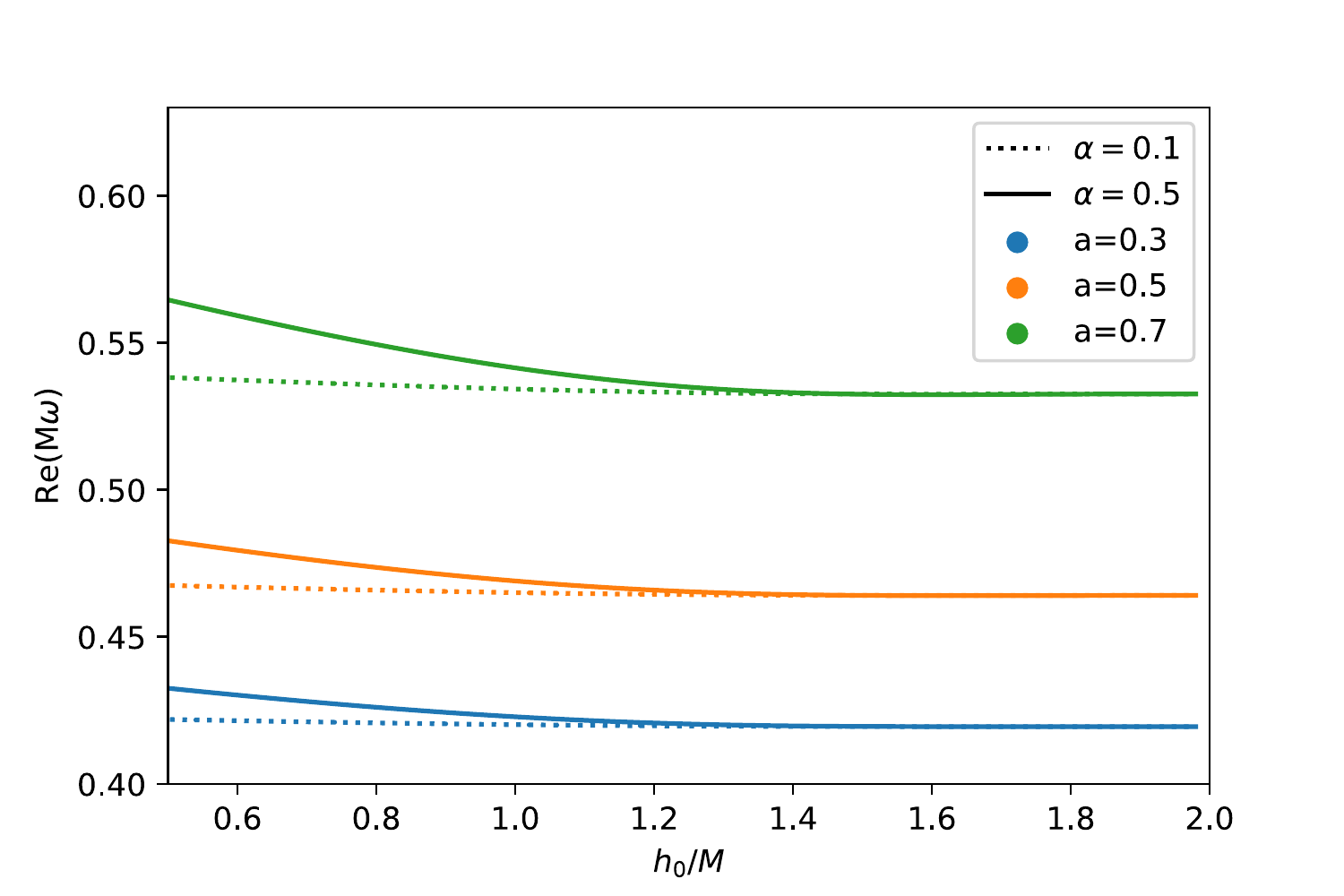}
   \includegraphics[scale=0.6]{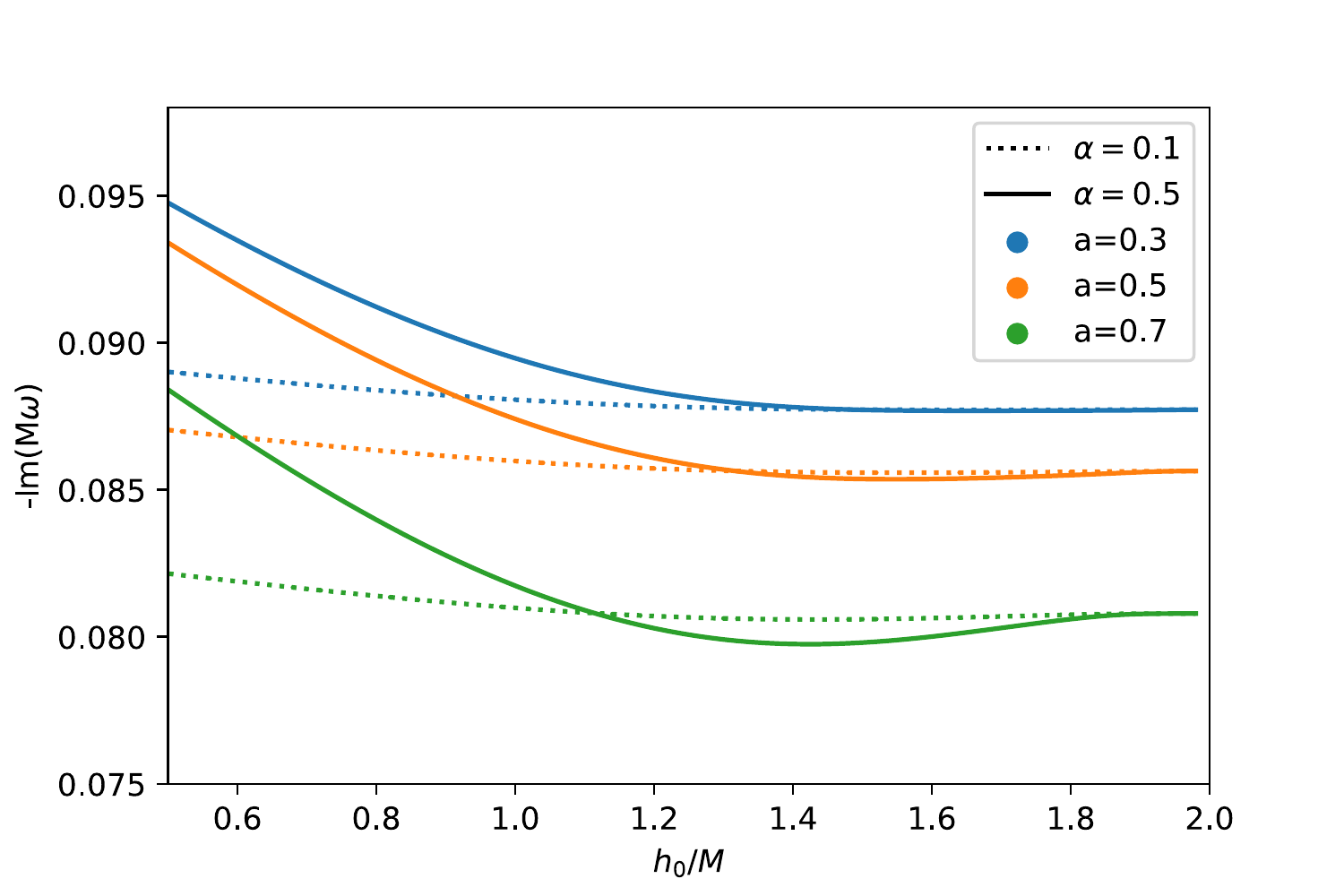}
  \caption{The $s=-2$ fundamental quasinormal modes with $l=m=2$ as a function of parameter $h_0\in[0.5,2]$, with different spin $a$ and $\alpha$. The upper panel is the real part while the low panel is the minus imaginary part. }
\label{reim-h}
\end{figure}

\begin{table*}[]
\setlength{\tabcolsep}{3mm}
\begin{tabular}{cllllllll}
\hline \hline
$h_0=1$ & \multicolumn{2}{c}{$\alpha$ = 0, Kerr}  & \multicolumn{2}{c}{$\alpha$=0.1} & \multicolumn{2}{c}{$\alpha$=0.5} & \multicolumn{2}{c}{$\alpha$=1}                             \\
$a$    & Re($\omega$) & -Im($\omega$) & Re($\omega$) & -Im($\omega$) & Re($\omega$) & -Im($\omega$) & Re($\omega$) & -Im($\omega$) \\ \hline
0.1 & 0.387018 & 0.088706 & 0.387514 & 0.089016 & 0.389570 & 0.090307 & 0.392310 & 0.092044 \\
0.2 & 0.402145 & 0.088311 & 0.402704 & 0.088634 & 0.405023 & 0.089980 & 0.408122 & 0.091795 \\
0.3 & 0.419527 & 0.087729 & 0.420173 & 0.088067 & 0.422858 & 0.089476 & 0.426465 & 0.091385 \\
0.4 & 0.439842 & 0.086882 & 0.440609 & 0.087229 & 0.443813 & 0.088683 & 0.448146 & 0.090663 \\
0.5 & 0.464123 & 0.085639 & 0.465067 & 0.085980 & 0.469026 & 0.087410 & 0.474434 & 0.089363 \\
0.6 & 0.494045 & 0.083765 & 0.495261 & 0.084067 & 0.500398 & 0.085320 & 0.507520 & 0.087007 \\
0.7 & 0.532600 & 0.080793 & 0.534283 & 0.080988 & 0.541479 & 0.081742 & 0.551710 & 0.082628 \\
0.8 & 0.586017 & 0.075630 & 0.588656 & 0.075563 & 0.600233 & 0.075057 & 0.617659 & 0.073752 \\
0.9 & 0.671614 & 0.064869 & 0.677072 & 0.064026 & 0.703157 & 0.059362 & 0.753330 & 0.048227

        \\ \hline\hline
\end{tabular}
\caption{Values of the quasinormal modes for the fundamental mode, with $s=-2$, $l =m=2$, $h_0=1$ with different values of $\alpha$ and spin $a$. $\alpha=0$ corresponds to Kerr black hole }
\label{tab1}
\end{table*}

\begin{table*}[]
\setlength{\tabcolsep}{3mm}
\begin{tabular}{cllllllll}
\hline \hline
$\alpha=0.5$ & \multicolumn{2}{c}{$h_0$ = 0.5}  & \multicolumn{2}{c}{$h_0$=1} & \multicolumn{2}{c}{$h_0$=1.5} & \multicolumn{2}{c}{$h_0$=1.99}                             \\
$a$    & Re($\omega$) & -Im($\omega$) & Re($\omega$) & -Im($\omega$) & Re($\omega$) & -Im($\omega$) & Re($\omega$) & -Im($\omega$) \\ \hline
0.1 & 0.397183 & 0.095178 & 0.389570 & 0.090307 & 0.387063 & 0.088734 & 0.387017 & 0.088706 \\
0.2 & 0.413525 & 0.095024 & 0.405023 & 0.089980 & 0.402195 & 0.088333 & 0.402145 & 0.088311 \\
0.3 & 0.432555 & 0.094764 & 0.422858 & 0.089476 & 0.419576 & 0.087720 & 0.419527 & 0.087729 \\
0.4 & 0.455153 & 0.094289 & 0.443813 & 0.088683 & 0.439877 & 0.086787 & 0.439842 & 0.086882 \\
0.5 & 0.482712 & 0.093406 & 0.469026 & 0.087410 & 0.464125 & 0.085370 & 0.464123 & 0.085639 \\
0.6 & 0.517632 & 0.091731 & 0.500398 & 0.085320 & 0.493995 & 0.083205 & 0.494045 & 0.083765 \\
0.7 & 0.564612 & 0.088411 & 0.541479 & 0.081742 & 0.532508 & 0.079807 & 0.532600 & 0.080793 \\
0.8 & 0.635179 & 0.081067 & 0.600233 & 0.075057 & 0.586002 & 0.074068 & 0.586017 & 0.075630 \\
0.9 & 0.778872 & 0.057647 & 0.703157 & 0.059362 & 0.672405 & 0.062476 & 0.671614 & 0.064869
\\ \hline\hline
\end{tabular}
\caption{Values of the quasinormal modes for the fundamental mode, with $s=-2$, $l =m = 2$, $\alpha=0.5$ with different values of $h_0$ and spin $a$.}
\label{tab2}
\end{table*}

\begin{table*}[]
\setlength{\tabcolsep}{3mm}
\begin{tabular}{cllllllll}
\hline \hline
$\alpha$=0.5, $h_0=1$ & \multicolumn{2}{c}{$n$ = 0 ($\alpha$=0)}  & \multicolumn{2}{c}{$n$=0} & \multicolumn{2}{c}{$n$=1} & \multicolumn{2}{c}{$n$=2}                             \\
$a$    & Re($\omega$) & -Im($\omega$) & Re($\omega$) & -Im($\omega$) & Re($\omega$) & -Im($\omega$) & Re($\omega$) & -Im($\omega$) \\ \hline
0.1 & 0.387018 & 0.088706  & 0.389570 & 0.090307 & 0.363668 & 0.275243 & 0.319704 & 0.477473 \\
0.2 & 0.402145 & 0.088311  & 0.405023 & 0.089980 & 0.381088 & 0.273410 & 0.340243 & 0.471936 \\
0.3 & 0.419527 & 0.087729  & 0.422858 & 0.089476 & 0.400989 & 0.270985 & 0.363435 & 0.465366 \\
0.4 & 0.439842 & 0.086882  & 0.443813 & 0.088683 & 0.424145 & 0.267668 & 0.390088 & 0.457279 \\
0.5 & 0.464123 & 0.085639  & 0.469026 & 0.087410 & 0.451751 & 0.262947 & 0.421462 & 0.446879 \\
0.6 & 0.494045 & 0.083765  & 0.500398 & 0.085320 & 0.485800 & 0.255900 & 0.459693 & 0.432749 \\
0.7 & 0.532600 & 0.080793 & 0.541479 & 0.081742 & 0.530002 & 0.244640 & 0.508836 & 0.411948 \\
0.8 & 0.586017 & 0.075630  & 0.600233 & 0.075057 & 0.592584 & 0.224438 & 0.577928 & 0.376740 \\
0.9 & 0.671614 & 0.064869  & 0.703157 & 0.059362 & 0.700220 & 0.177572 & 0.694559 & 0.296919
\\\hline\hline
\end{tabular}
\caption{Values of the quasinormal modes for the fundamental mode $n=0$ and few overtones $n=$1, 2, with $s=-2$, $l =m = 2$, $\alpha=0.5$, $h_0=1$ for different values of spin $a$. The $n$ = 0 ($\alpha$=0) column correspond to the Kerr black hole case. As one expect, the higher overtone, the smaller real part and bigger imaginary part of quasinormal modes. }
\label{tab3}
\end{table*}

To better show the dependence of the quasinormal frequencies on the quantum number $(l,m)$ and the spin of the black hole, we have also computed ($s=-2$, $l=2,3$, $m=-l....l$) quasinormal frequencies from $a=0.01$ to $a=0.91$, with parameters $\alpha=0.5$, $h_0=1$, The results are shown in Fig.\ref{lm}. Please note that here and in all further figures we have used the minus $Im(\omega)$ for the imaginary part of the quasinormal modes. We see that the quantum number $l$ causes only a general shift of the quasinormal modes. However, non-negative $m$ causes the real part of the quasinormal modes to increase with spin (larger $m$, larger ratio of increase), while negative $m$ causes the opposite.
For the imaginary part, all decrease with spin.

Next, we want to investigate how the hairy parameters $\alpha$ and $h_0$ affect the relation between the quasinormal modes and the spin of the black hole. Therefore, in Fig.\ref{reim-a} we plot the real and imaginary parts of the quasinormal frequencies as a function of the black hole spin, with different combinations of the parameters $\alpha$ and $h_0$. We can see that $\alpha$ and $h_0$
do not affect the overall tendency between quasinormal modes and the spin of the black hole, but slightly change the slopes. In particular, for the imaginary part, there are crossings between different combinations of the $\alpha$ and $h_0$ parameters in the high spin regime.

To see it more clearly, we examine respectively how $\alpha$ and $h_0$ will affect the quasinormal frequencies. We have plotted the ($s=-2$, $l=m=2$) fundamental quasinormal modes as a function of the parameter $\alpha$ in Fig.\ref{reim-alpha} and $h_0$ in Fig.\ref{reim-h}, by fixing three spin and the other one parameter. We find that both the real and imaginary parts of the quasinormal modes increase monotonically with $\alpha$. However, the reverse is true for $h_0$: the larger $h_0$ becomes, the smaller the quasinormal modes become, and they almost become a constant when they reach the limit $h_0=h_K$. Nevertheless, we can see from both Fig.\ref{reim-alpha} and Fig.\ref{reim-h} that the larger the spin, the larger the real part of the quasinormal modes and the smaller the imaginary part.

\section{conclusion and discussion}\label{sec6}
In this work, we study the phenomenology of gravitational perturbation around rotating hairy black holes. We first introduced the hairy black hole (\ref{metric}) and also used an approximation method to obtain the horizons analytically. Then we derived the master equations of the gravitational perturbation field including a radial part (\ref{radial}) and an angular part (\ref{angular}). Based on these equations, we studied the superradiance instability and quasinormal modes of the rotating hairy black holes.

For superradiance, we derived the conditions for superradiance to occur. The results show that amplification occurs when certain frequency criteria are met (\ref{cds}). Then, we calculated the amplification factor with low frequency approximations by using the matching-asymptotic technique. In the end, we obtained the formula to calculate amplification factor (\ref{ll}) and plotted the $s=-2,l=m=2$ modes with different parameters.

As for the quasinormal modes, we have calculated the quasinormal modes of the rotating hairy black hole numerically using the continued fraction method. We first present three tables (Table.\ref{tab1}, \ref{tab2}, \ref{tab3}), and the columns $\alpha=0$ in Table.\ref{tab1} and Table.\ref{tab3} could validate our numerical approaches. They agree very well with the previous results in the Kerr limit \cite{valid}. Then we demonstrated and drew several figures to show how the quasinormal modes change with the variation of different parameters: (a) fundamental quasinormal modes as a function of quantum number $(l,m)$; (b) fundamental quasinormal modes as a function of black hole spin $a$; (c) fundamental quasinormal modes as a function of hairy parameters $\alpha, h_0$; By fixing the other parameters respectively.

The results of both superradiance instability and quasinormal modes show that the rotating hairy black hole is different from a Kerr black hole. Although the discrepancies are small, one might still expect that this should have some consequences for observations such as GW, in particular the discovery and accurate identification of the ringdown signal could provide the 'smoking gun' for testing this rotating hairy black hole. We will continue these studies in the future.

\section*{acknowledgments}
The author would like to thank Professor Steen Hansen for helpful discussions. The author also thanks the DARK Cosmology Centre at the Niels Bohr Institute for support of this research. This work was also financially supported by the China Scholarship Council.
\\
\\
\begin{appendix}
\section{THE ERRORS OF APPROXIMATION METHOD}\label{ap1}

\begin{figure*}[htbp]
  \includegraphics[scale=0.58]{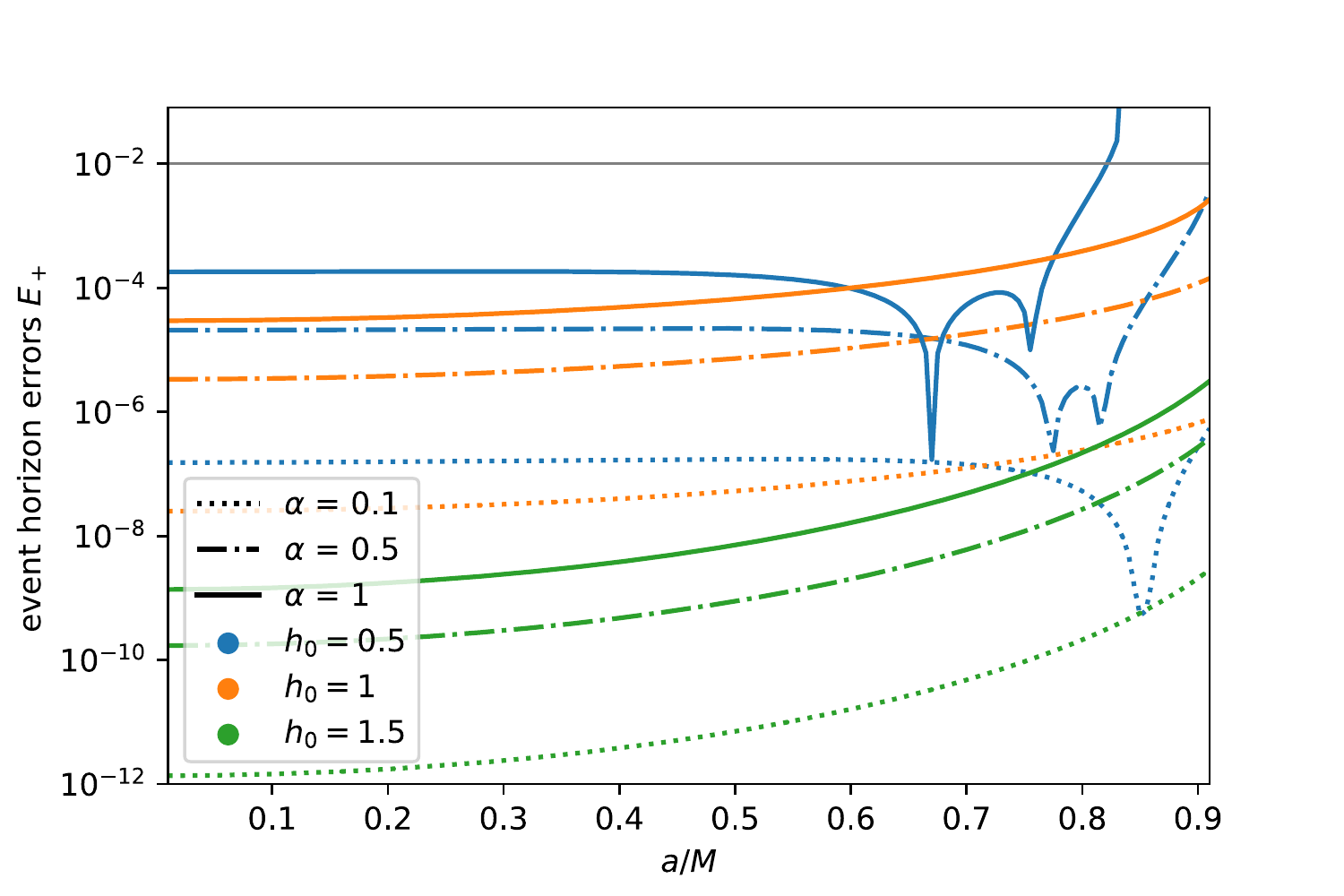}
   \includegraphics[scale=0.58]{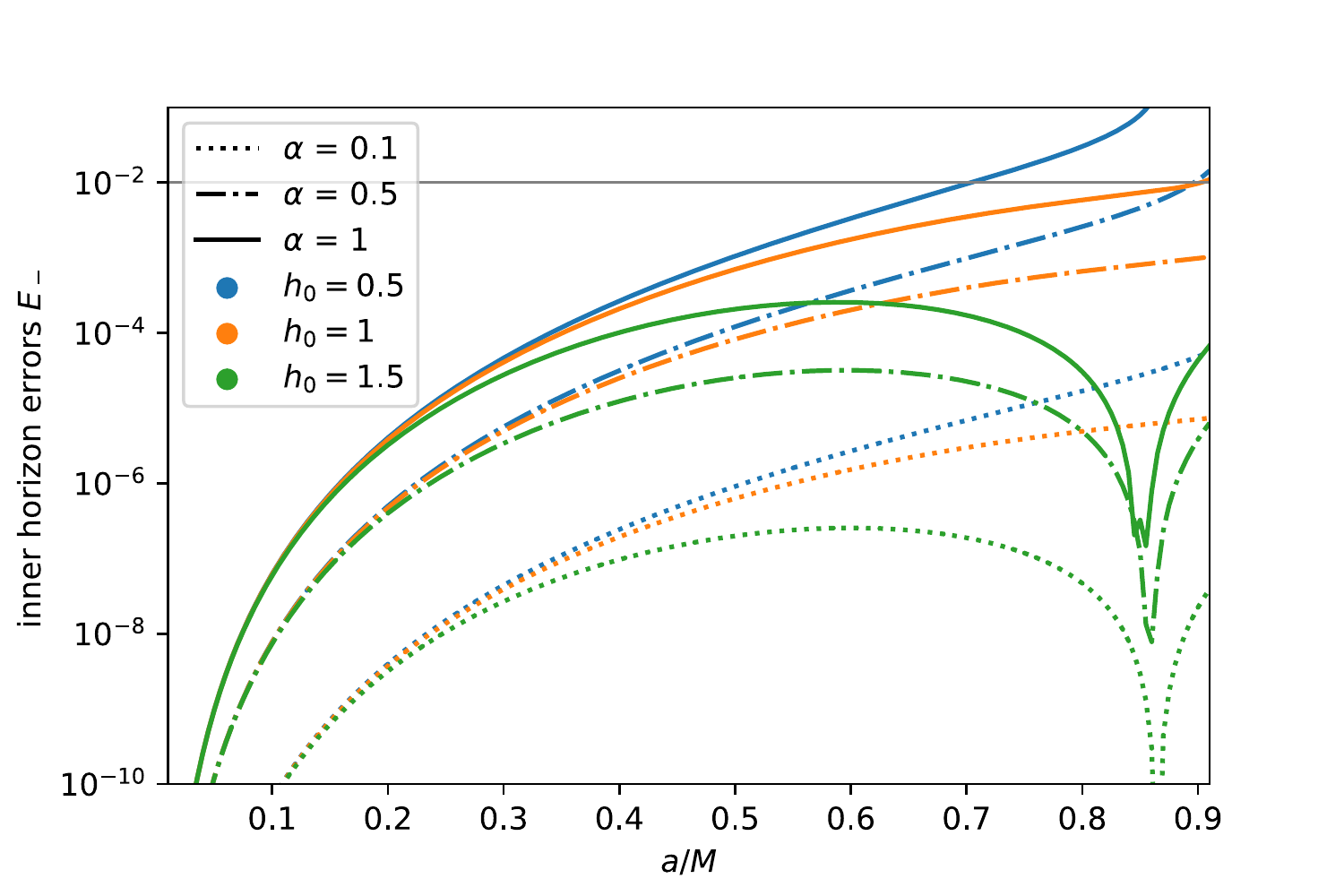}
\caption{Upper and lower plots are respectively the errors of second order event horizon and inner horizon, with parameter $\alpha=0.1,0.5,1$, $l_0=0.5, 1, 1.5$ and black hole spin $a=0.01$ to $0.91$.}
\label{err}
\end{figure*}

In Sec.\ref{sec2}, we used the analytical approximation to solve the Delta function (\ref{eh}). To show the accuracy of this method, we plot the errors of the second order solutions in Fig.\ref{err} as a function of the spin $a$ of the black hole at different parameters $\alpha, h_0$. The errors, denoted as $E_{\pm}$, are calculated by comparison with the numerical solutions $r_{\pm}^{num}$ of (\ref{eh}), i.e,
\begin{equation}
E_{\pm}=\frac{\left| \hat r_{\pm} - r_{\pm}^{num}\right|}{r_{\pm}^{num} }
\end{equation}

We can see from Fig.\ref{err}, the approximation errors for both the event horizon and the inner horizon increase with spin. There is some oscillation for the event horizon errors as the black hole spin approaches higher values. We want to control the errors so that they never increase above $10^{-2}$ within the spin intervals $a=0.01$ to $0.9$. Therefore, we draw a horizontal line $E_{\pm}=10^{-2}$ in Fig.\ref{err}. Then the parameters $\alpha, h_0, a$ should be chosen within a certain range so that the approximation works well, and for this purpose second order solutions are sufficient.

\begin{figure}[htbp]
\centering
\includegraphics[scale=0.9]{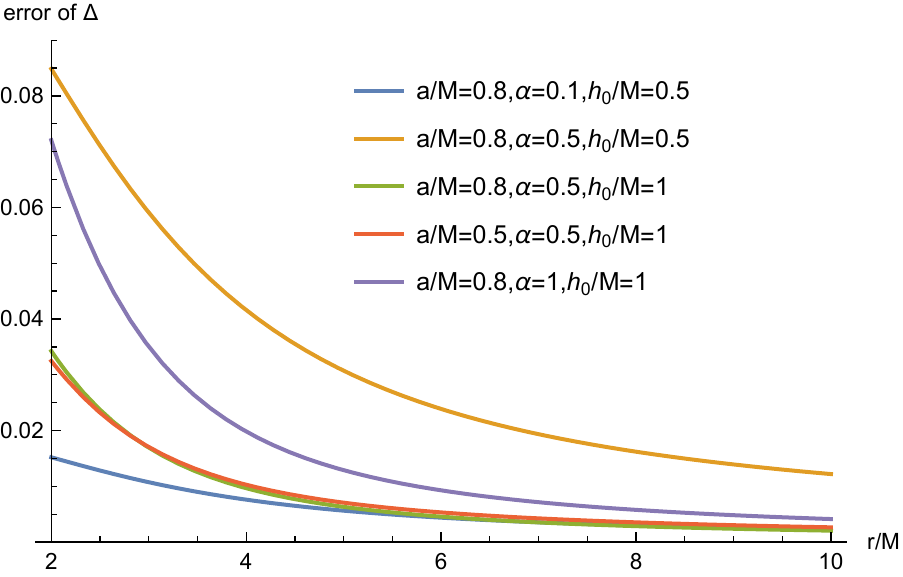}
\caption{The relative error of Delta as a function of radial distance. We plot the errors with several combinations of black spin $a$, $\alpha$ and $h_0$ as references. These combination values were used in this work. }
\label{ed}
\end{figure}

It is true that the original Delta function depends not only on the roots, but everywhere in the radial distance. However, replacing the original Delta function (which is a transcendental function) with a quadratic function gives a very good approximation for $r\gtrsim 2M\approx \hat r_+$ (see Fig.\ref{ed}), i.e., outside the event horizon, since all perturbations occur in the region $r\gtrsim 2M$. The errors are calculated by
\begin{equation}
\frac{\Delta_{approx}-\Delta_{original}}{\Delta_{original}}
\end{equation}
where $\Delta_{approx}$ is given by Eq.(\ref{eh2}) and $\Delta_{original}$ is the exact Delta function in metric (\ref{metric}). We can clearly see that the errors decrease rapidly as r becomes large. The larger $\alpha$ and smaller $h_0$, the larger the errors. The spin has only a small effect. The errors go below a few percent when $r\gtrsim 2M$. One can also notice that the $\Delta_{approx}$ is always slightly larger than the $\Delta_{original}$.

For a quadratic function Eq.(\ref{eh2}), the roots determine its behavior everywhere outside the event horizon. Moreover, the final equation for calculating the amplification factor and quasinormal modes are (\ref{ll}) and (\ref{cf}) respectively, the essential quantities $Q$, $k$, $\alpha_n, \beta_n, \gamma_n$ for these equations are all proportional to $\hat r_+$, $\hat r_-$ or their quadratic. Therefore, the accuracy of amplification factor and quasinormal modes, because of using the approximation solutions (\ref{r++}) and (\ref{r--}), are therefore in the order of $(E_{\pm})^{2l+2}$ and $E_{\pm}$, respectively. Choosing the parameters within a certain range $\alpha \in [0,0.5], h_0 \in [0.5,2], a \in[0.01,0.9]$, the errors of amplification factor and quasinormal modes are always below the order of $10^{-4l+4}$ and $10^{-2}$, respectively.
\end{appendix}


\begin{thebibliography}{0}%
\makeatletter
\providecommand \@ifxundefined [1]{%
 \@ifx{#1\undefined}
}%
\providecommand \@ifnum [1]{%
 \ifnum #1\expandafter \@firstoftwo
 \else \expandafter \@secondoftwo
 \fi
}%
\providecommand \@ifx [1]{%
 \ifx #1\expandafter \@firstoftwo
 \else \expandafter \@secondoftwo
 \fi
}%
\providecommand \natexlab [1]{#1}%
\providecommand \enquote  [1]{``#1''}%
\providecommand \bibnamefont  [1]{#1}%
\providecommand \bibfnamefont [1]{#1}%
\providecommand \citenamefont [1]{#1}%
\providecommand \href@noop [0]{\@secondoftwo}%
\providecommand \href [0]{\begingroup \@sanitize@url \@href}%
\providecommand \@href[1]{\@@startlink{#1}\@@href}%
\providecommand \@@href[1]{\endgroup#1\@@endlink}%
\providecommand \@sanitize@url [0]{\catcode `\\12\catcode `\$12\catcode
  `\&12\catcode `\#12\catcode `\^12\catcode `\_12\catcode `\%12\relax}%
\providecommand \@@startlink[1]{}%
\providecommand \@@endlink[0]{}%
\providecommand \url  [0]{\begingroup\@sanitize@url \@url }%
\providecommand \@url [1]{\endgroup\@href {#1}{\urlprefix }}%
\providecommand \urlprefix  [0]{URL }%
\providecommand \Eprint [0]{\href }%
\providecommand \doibase [0]{https://doi.org/}%
\providecommand \selectlanguage [0]{\@gobble}%
\providecommand \bibinfo  [0]{\@secondoftwo}%
\providecommand \bibfield  [0]{\@secondoftwo}%
\providecommand \translation [1]{[#1]}%
\providecommand \BibitemOpen [0]{}%
\providecommand \bibitemStop [0]{}%
\providecommand \bibitemNoStop [0]{.\EOS\space}%
\providecommand \EOS [0]{\spacefactor3000\relax}%
\providecommand \BibitemShut  [1]{\csname bibitem#1\endcsname}%
\let\auto@bib@innerbib\@empty
\end{thebibliography}%


\begin{thebibliography}{100}
\bibitem{gw1}
B. P. Abbott et al. (LIGO Scientific, Virgo Collaborations), Phys. Rev. Lett. {\bf 116}, 061102 (2016)
\bibitem{gw2}
B. P. Abbott et al. (LIGO Scientific, Virgo Collaborations), Phys. Rev. D. {\bf 100}, 104036 (2019)
\bibitem{gw3}
B. P. Abbott et al. (LIGO Scientific, Virgo Collaborations), Phys. Rev. Lett. {\bf 119}, 161101 (2017)
\bibitem{shadow1}
K. Akiyama et al. (Event Horizon Telescope Collaborations), Astrophys. J. Lett. {\bf 875}, L1 (2019)
\bibitem{shadow2}
K. Akiyama et al. (Event Horizon Telescope Collaborations), Astrophys. J. Lett. {\bf 930}, L15
\bibitem{nh1}
B. Carter. Phys. Rev. Lett. {\bf 26}, 331 (1971).
\bibitem{nh2}
S. W. Hawking. Communications in Mathematical Physics. {\bf 25}, 152 (1972).
\bibitem{nh3}
D. C. Robinson. Phys. Rev. Lett. {\bf 34}, 905 (1975). 
\bibitem{nh4}
P. O. Mazur. Journal of Physics A Mathematical General. {\bf 15}, 3173 (1982). 
\bibitem{nh5}
J. D. Bekenstein. Phys. Rev. D.  {\bf 51}, 6608 (1995).
\bibitem{dk}
C. A. R. Herdeiro, E. Radu. Int. J. Mod. Phys. D, {\bf 24},  1542014 (2015).
\bibitem{gd0}
J. Ovalle, R. Casadio, E. Contreras and A. Sotomayor. Phys. Dark Univ. {\bf 31}, 100744 (2021).
\bibitem{gd}
E. Contreras, J. Ovalle, R. Casadio. Phys. Rev. D, {\bf 103}, 044020 (2021).
\bibitem{gd1}
J. Ovalle. Phys. Rev. D {\bf 95}, 104019 (2017).
\bibitem{gd2}
J. Ovalle. Phys. Lett. B {\bf 788}, 213 (2019).
\bibitem{inv1}
S. U. Islam, S. G. Ghosh. Phys. Rev. D {\bf 103} 124052 (2021).
\bibitem{inv2}
M. Afrin, R. Kumar and S. G. Ghosh. Mon. Not. Roy. Astron. Soc. {\bf 504}, 5927 (2021).
\bibitem{inv3}
R. T. Cavalcanti, R. C. de Paiva and R. da Rocha. Eur. Phys. J. Plus {\bf 137}, 1185 (2022).
\bibitem{inv31}
S. Vagnozzi, et al. [arXiv:2205.07787].
\bibitem{inv4}
S. K. Jha and A. Rahaman. [arXiv:2205.06052].
\bibitem{inv5}
S. Mahapatra and I. Banerjee. [arXiv:2208.05796].
\bibitem{rd}
J. Abadie, et al. Phys. Rev. D, {\bf 83}, 122005 (2011).
\bibitem{rd2}
O. Dreyer, et al. Class. Quantum Grav. {\bf 21}, 787 (2004).
\bibitem{sr}
R. Brito, V. Cardoso, and P. Pani, Lect. Notes Phys. {\bf 906}, 237 (2015).
\bibitem{srs}
R. Roy, S. Vagnozzi and L. Visinelli. Phys. Rev. D {\bf 105}, 083002 (2022).
\bibitem{srs1}
Y. Chen, R. Roy, S. Vagnozzi and L. Visinelli. Phys. Rev. D {\bf 106}, 043021 (2022).
\bibitem{sr2}
R. C. F. Hugo, A. R. H. Carlos. Phys. Rev. D {\bf 97}, 084003 (2018).
\bibitem{sr3}
T. Kolyvaris, M. Koukouvaou, A. Machattou and E. Papantonopoulos. Phys. Rev. D {\bf 98}, 024045 (2018).
\bibitem{sr4}
V. P. Frolov and A. Zelnikov. Phys. Rev. D {\bf 98}, 084035 (2018).
\bibitem{sr5}
M. F. Wondrak, P. Nicolini and J. W. Moffat. JCAP {\bf 12}, 021 (2018).
\bibitem{sr6}
K. Destounis. Phys. Rev. D {\bf 100}, 044054 (2019).
\bibitem{sr7}
M. Khodadi, A. Talebian and H. Firouzjahi, [arXiv:2002.10496].
\bibitem{sr8}
E. Franzin, S. Liberati and M. Oi. Phys.Rev.D {\bf 103}, 104034 (2021).
\bibitem{sr81}
E. Franzin, S. Liberati, J. Mazza, R. Dey and S. Chakraborty. Phys. Rev. D {\bf 105}, 124051 (2022).
\bibitem{sr82}
R. Dey, S. Biswas and S. Chakraborty. Phys. Rev. D {\bf 103}, 084019 (2021).
\bibitem{sr9}
M. Khodadi. Phys. Rev. D {\bf 103}, 064051 (2021).
\bibitem{sr10}
W. X. Chen and Y. G. Zheng. [arXiv:2103.04239].
\bibitem{sr11}
B. Cuadros-Melgar, R. D. B. Fontana and J. de Oliveira. Phys. Rev. D {\bf 104}, 104039 (2021).
\bibitem{sr12}
M. Khodadi and R. Pourkhodabakhshi. Phys. Lett. B {\bf 823}, 136775 (2021).
\bibitem{sr13}
M. G. Richarte, \'E. L. Martins and J. C. Fabris. Phys. Rev. D {\bf 105}, 064043 (2022).
\bibitem{sr14}
S. Alexander, G. Gabadadze, L. Jenks and N. Yunes. [arXiv:2201.02220].
\bibitem{sr15}
G. Mascher, K. Destounis, K. D. Kokkotas. Phys. Rev. D {\bf 105}, 084052 (2022).
\bibitem{sr16}
T. Ishii, Y. Kaku and K. Murata. JHEP {\bf10}, 024 (2022).
\bibitem{sr17}
S. K. Jha and A. Rahaman. [arXiv:2208.13176].
\bibitem{sr18}
H. Yang and Y. G. Miao. [arXiv:2211.15130].
\bibitem{sr19}
Zeldovich, JETP Lett. {\bf 14}, 180 (1971); JETP {\bf35}, 1085 (1972);
\bibitem{sr20}
T.~Torres, etc, Nature Phys. {\bf13}, 833 (2017).
\bibitem{sr21}
M.~C.~Braidotti,etc, Phys. Rev. Lett. {\bf 128}, 013901 (2022).
\bibitem{zhen}
Zhen Li, Phys.Rev.D {\bf107} 044013 (2023). [arXiv:2210.14062].
\bibitem{nre2}
S. G. Ghosh, Eur. Phys. J. C. {\bf 75}, 532 (2015).
\bibitem{tes}
S. A. Teukolsky, Astrophys. J. {\bf 185}, 635 (1973).
\bibitem{df}
A.~L.~Dudley and J.~D.~Finley, III, J. Math. Phys. \textbf{20} (1979), 311.
\bibitem{kk}
K. D. Kokkotas, Nuov Cim B {\bf 108}, 991 (1993).
\bibitem{kk1}
E.~Berti and K.~D.~Kokkotas, Phys. Rev. D \textbf{71} (2005), 124008.
\bibitem{extr}
R.~Brito, V.~Cardoso and P.~Pani, Class. Quant. Grav. {\bf 32}, 134001 2015.
\bibitem{asf}
Handbook of Mathematical Functions with Formulas, Graphs, and Mathematical Tables, edited by M. Abramowitz and I. A. Stegun (Dover, New York, 1965).
\bibitem{s-s}
D. N. Page, Phys.Rev. D. {\bf 13}, 198 (1976).
\bibitem{match}
A. Starobinski, Zh. Eksp. Teor. Fiz. {\bf 64}, 48 (1973).
\bibitem{match1}
A. Starobinski and S. M. Churilov, Zh. Eksp. Teor. Fiz. {\bf 65}, 3 (1973).
\bibitem{match2}
D.~N.~Page, Phys. Rev. D {\bf 13}, 198 (1976).
\bibitem{qnm}
K. D. Kokkotas and B. G. Schmidt, Living Rev. Rel. {\bf 2}, 2 (1999).
\bibitem{qnm1}
E. Berti, V. Cardoso and A. O. Starinets, Class. Quant. Grav. {\bf 26}, 163001 (2009).
\bibitem{qnm2}
R. A. Konoplya and A.V. Zhidenko, Rev. Mod. Phys. {\bf 83}, 793 (2011).
\bibitem{cfm}
P. H. C. Siqueira and M. Richartz,
Phys. Rev. D. {\bf 106}, 024046 (2022).
\bibitem{cfm1}
A. K. Mishra, A. Ghosh and S. Chakraborty, Eur. Phys. J. C. {\bf 82}, 820 (2022).
\bibitem{cfm2}
R. G. Daghigh, M. D. Green and J. C. Morey, [arXiv:2209.09324].
\bibitem{lea}
E. W. Leaver, Proc. Roy. Soc. Lond. A {\bf 402}, 285 (1985).
\bibitem{nl}
H. P. Nollert, Phys. Rev. D. {\bf 47}, 5253 (1993).
\bibitem{valid}
E. Berti, V. Cardoso and C. M. Will. Phys. Rev. D {\bf 73}, 064030 (2006).
\bibitem{valid1}
R. A. Konoplya and A. V. Zhidenko, Phys. Rev. D. {\bf 73}, 124040 (2006).
\bibitem{valid2}
S. R. Dolan, Phys. Rev. D. {\bf 76}, 084001 (2007).
\end{thebibliography}
\end{document}